\documentclass[osajnl,showpacs,twocolumn]{revtex4}

\usepackage{hyperref} 

\usepackage{amsmath}
\usepackage{xspace}

\newcommand{\mm}{\thinspace \ensuremath{\mathrm{mm}}\xspace}

\newcommand{\micron}{\thinspace\ensuremath{\mu \mathrm{m}}\xspace}
\newcommand{\nm}{\thinspace \ensuremath{\mathrm{nm}}\xspace}

\newcommand{\invnm}{\thinspace\ensuremath{\mathrm{nm}^{-1}}\xspace}
\newcommand{\s}{\thinspace\ensuremath{\mathrm{s}}\xspace}
\newcommand{\ms}{\thinspace\ensuremath{\mathrm{ms}}\xspace}

\newcommand{\mA}{\thinspace \ensuremath{\mathrm{mA}}\xspace}

\newcommand{\degree}{\ensuremath{^\circ}\xspace}
\newcommand{\cc}[1]{{#1}^{\ast}}                            

\newcommand{\bvec}[1]{\ensuremath{\mathbf{#1}}}
\newcommand{\GB}{\thinspace \ensuremath{\mathrm{GB}}\xspace}
\newcommand{\MB}{\thinspace \ensuremath{\mathrm{MB}}\xspace}

\begin{document}

\title{High-resolution \emph{ab initio} three-dimensional X-ray diffraction microscopy}
\author{H.~N.~Chapman}\email{henry.chapman@llnl.gov}
\author{A.~Barty}
\author{S.~Marchesini}
\author{A.~Noy}
\affiliation{University of California, Lawrence Livermore National
  Laboratory, 7000 East Ave., Livermore, CA 94550, USA}
\author{C.~Cui, M.~R.~Howells, R.~Rosen}
\affiliation{Advanced Light Source, Lawrence Berkeley National Laboratory,
  1 Cyclotron Road, Berkeley, CA 94720, USA}
\author{H.~He, J.~C.~H.~Spence, U.~Weierstall}
\affiliation{Department of Physics and Astronomy, Arizona State University,
  Tempe, AZ 85287-1504, USA}
\author{T.~Beetz, C.~Jacobsen, and D.~Shapiro}
\affiliation{Department of Physics and Astronomy, Stony Brook University,
  Stony Brook, NY 11794-3800, USA}

\begin{abstract}
  Coherent X-ray diffraction microscopy is a method of imaging
  non-periodic isolated objects at resolutions only limited, in
  principle, by the largest scattering angles recorded.  We
  demonstrate X-ray diffraction imaging with high resolution in all
  three dimensions, as determined by a quantitative analysis of the
  reconstructed volume images.  These images are retrieved from the 3D
  diffraction data using no \emph{a priori} knowledge about the shape
  or composition of the object, which has never before been
  demonstrated on a non-periodic object. We also construct 2D images
  of thick objects with infinite depth of focus (without loss of
  transverse spatial resolution).  These methods can be used to image
  biological and materials science samples at high resolution using
  X-ray undulator radiation, and establishes the techniques to be used
  in atomic-resolution ultrafast imaging at X-ray free-electron laser
  sources.
\end{abstract}

\pacs{340.7460, 110.1650, 110.6880, 100.5070, 100.6890, 070.2590, 180.6900}

\maketitle

\section{Introduction}
\label{sec:intro}
In many fields of science the ability to visualize the
three-dimensional organization of component parts is proving crucial
to our understanding of the mechanisms involved in atomic and
molecular processes. This is occurring in fields as diverse as
whole-cell imaging in biology, the study of the minimum energy pathway
for crack-propagation in brittle solids, and the internal structure of
the new labyrinthine mesoporous structures developed by inorganic
chemists for a wide range of applications.

The field of coherent X-ray diffraction imaging (CXDI, also known as
diffraction microscopy) is expected to make a significant contribution
to this effort.  In this method, first put forward and developed by
David Sayre \cite{Sayre:1995, Sayre:1998}, an image is reconstructed
from measurements of the far-field scattered intensity of an isolated
and non-periodic object.  The resolution of this form of microscopy is
limited only by the wavelength and the largest scattering angle
recorded.  Hence this method is being pursued as a method for
high-resolution X-ray microscopy without the technological limitations
of manufacturing high-resolution optical elements \cite{Miao:1999,
Robinson:2001, Williams:2003, Marchesini:2003, Marchesini:2003a,
He:2003} The penetrating nature of X-rays allows imaging of objects
much thicker than can be examined in a TEM (e.g. 10\micron), at
resolutions much better than visible microscopes.  Preliminary studies
of radiation damage suggest that 3D resolutions of about 10\nm should
be achievable on frozen hydrated biological material
\cite{Howells:2005}.  The method is also being pursued in order to
push X-ray imaging to its resolution limits, namely ultrafast
near-atomic-resolution imaging of macromolecules at X-ray
free-electron laser (XFEL) sources \cite{Neutze:2000, Miao:2001} and
of laser-aligned molecules \cite{Spence:2004}, that will enable
structure determination without the need for crystallizing material.

High resolution imaging of thick objects can only be attained in the
context of three-dimensional (3D) measurement and reconstruction.  In
most cases, other than surface studies or imaging of man-made objects,
the analysis of the structure can only be properly interpreted in
three dimensions.  Unless the object itself is a slice of material
that is thinner than the depth of focus of a two-dimensional (2D)
image, artifact-free structural analysis can only be carried out with
knowledge of the surrounding material, or by applying imaging
modalities whereby depth information is not strongly transferred to
the image (such as confocal imaging).  At resolution lengths very much
larger than the wavelength, thickness effects do not play a
significant role since, at the correspondingly low numerical aperture,
the depth of focus may be much larger than the size of the object.
This is certainly the case as one satisfies the projection
approximation of high energy X-ray tomography, where the depth of
focus approaches infinity.  Tomographic imaging in this mode is
limited by detector pixel size, or, if a diverging beam is used, by
Fresnel diffraction effects.  However, as one moves to higher
resolution, the depth of focus decreases dramatically, with the ratio
of transverse resolution length to longitudinal depth of focus given
by the numerical aperture.  For the classes of high-resolution
microscopy and structure determination applications in which we are
interested that imaging can only properly be carried out in 3D.

Coherent 3D X-ray diffraction imaging is especially suited to
high-resolution X-ray microscopy.
With a collimated beam incident on an object, the far-field
diffraction pattern (recorded on a flat CCD) represents diffracted
intensities which, in reciprocal space, sample points on the momentum-
and energy-conserving Ewald sphere. By rotating the sample about an
axis normal to the beam, this sphere, which passes through the origin,
sweeps through almost all of the reciprocal space volume of continuous
diffuse scattering from our non-periodic object. In this way we
collect the three-dimensional distribution of scattered intensity in
reciprocal space, which is phased using the 3D implementations of
iterative methods, as discussed below.  Once the phases of the
diffraction intensities in the diffraction volume have been
determined, the 3D Fourier transform of the object is known and the 3D
image can be obtained simply by an inverse Fourier transform. As will
be demonstrated in this paper, such datasets can be used for
artifact-free analysis of structures.  This is also the case for
crystallography, but is not generally the case for imaging with a
lens.  Partially-coherent tomographic imaging techniques, such as
tomography in the scanning transmission X-ray microscope (STXM)
\cite{Haddad:1994} or transmission X-ray microscope (TXM) \cite{Weiss:2000,
Larabell:2004}, lead to a complicated transfer of object spatial
frequencies into the measured image and there is no longer a simple
one-to-one mapping of a measurement on a detector pixel, for example,
to a spatial frequency of the object.  For some classes of object,
such as pure phase or amplitude objects, it may be possible to
deconvolve the 3D transfer function, but this is not generally assured
\cite{Streibl:1985}.  As with coherent diffraction imaging and
crystallography, coherent imaging with a lens also leads to a direct
mapping of spatial frequencies in the object to spatial frequencies of
the image.  Again, a tomographic reconstruction from coherent 2D
images can be easily achieved for pure amplitude or phase objects, but
would otherwise require knowing the phase and amplitude of the image
in order to transform into 3D reciprocal space. Coherent diffraction
imaging essentially attempts to emulate coherent lens-based imaging,
using a computer algorithm in place of a lens.  The advantage, for
tomography of complex objects, is that the diffraction amplitudes are
measured and the phases retrieved from the oversampling of those
amplitudes, so that a direct 3D Fourier synthesis of the object can be
achieved.

In this paper we perform an important demonstration of the feasibility
of high-resolution diffraction microscopy required for biological and
materials characterization, as well as single-molecule imaging.
Significantly this is done without the use of detailed \emph{a priori}
information about the sample structure or low-resolution data obtained
by other means. We also demonstrate that a full 3D reconstruction can
be produced on a $1024^3$ or larger data cube in a reasonable amount
of time using currently available computational hardware.  Three
significant recent developments have enabled us to perform full 3D
image reconstructions with high resolution in all three dimensions.
The commissioning of a new diffraction tomography apparatus by Stony
Brook University at an undulator beamline of the Advanced Light Source
(ALS) \cite{Beetz:2005} allows us to acquire diffraction patterns at
over one hundred orientations of an object, with short exposure times,
over angular ranges of more than $\pm 70\degree$.  The Shrinkwrap
phase-retrieval algorithm that we developed \cite{Marchesini:2003} has
proven to be extremely robust and effective in performing phase
retrieval on diffraction datasets with missing data (e.g. due to a
beam-stop) or limited angles.  The algorithm retrieves images \emph{ab
initio} from the measured diffraction intensities. It does not require
additional information about the object, such as a low-resolution
image, and can retrieve phases of general complex-valued objects.  The
third advance is the ability to perform 3D fast Fourier transforms
(FFTs) on the large $1024^3$-element arrays of diffraction data that
are assembled from our measurements.  Although the familiar increase
of computer processing power has brought giga-element FFTs in reach of
today's computers, it has been the development of computer clusters
and specific software for distributed computation of FFTs that has
made feasible the 3D implementation of the Shrinkwrap algorithm.  In
particular, we utilize the
\texttt{dist\_fft} software \cite{Crandall:2004} on a 16-node cluster of
dual-processor Apple G5 Xserves, giving us a performance of 8.6\s per
$1024^3$-element single-precision complex FFT.  We note that this
computational advance should also
benefit the field of diffraction tomography
\cite[Sec.~13.2]{Born:2002}, in which both the phase 
and amplitude of the scattered field are measured as is possible with
scattered ultrasonic waves.

We present here experimental results of high-resolution 3D X-ray
diffraction imaging of a well-characterized test object to demonstrate
the practical application of these advances and quantitatively assess
the technique.  We show the first full 3D X-ray diffraction images
that have been reconstructed without prior knowledge of the sample.
We believe that these are the highest resolution 3D X-ray images of
non-crystalline objects ever achieved, with a demonstrable resolution
volume of $10\nm
\times 10\nm \times 40\nm$.  We demonstrate that artifact-free 2D
images can be created from the 3D diffraction dataset of objects very
much thicker than the depth of focus.  In Sec.~\ref{sec:imaging} we
review diffraction imaging, the experimental requirements for 3D image
reconstructions, and our computer implementation to perform the 3D
phase retrieval and Fourier synthesis of the image.  Our sample
preparation and characterization techniques are discussed in
Sec.~\ref{sec:samples}, and our particular experimental setup and
methods are described in Secs.~\ref{sec:acquisition} and
\ref{sec:spectrogram}.  Image reconstruction results are presented in
Sec.~\ref{sec:reconstruction}. The 3D images are visualized as
iso-surface renderings, infinite depth-of-focus projection images,
maximum value projections, and tomographic slices through the
object. We also compare artifact-free 2D projections of 3D data to
reconstructions of individual 2D views, and illustrate the artifacts
present in single-view 2D images of thick objects.  In
Sec.\ref{sec:analysis} we quantitatively assess our 3D image
resolution.

\section{Three-Dimensional Coherent Diffraction Imaging}
\label{sec:imaging}
\begin{figure}[htbp]
  \centering
  \includegraphics[width=0.45\textwidth]{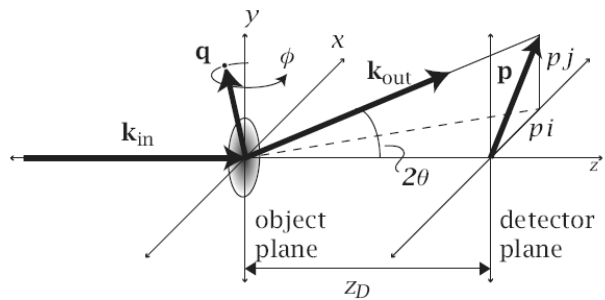}
  \caption{Scattering geometry for coherent X-ray diffraction imaging.  The sample is
    rotated about the $y$ axis by an angle $\phi$.}
  \label{fig:geometry}
\end{figure}

The incident X-ray wavefield
interacts with a three-dimensional (3D) periodic or non-period object through the
scattering potential of the object, $o(\bvec{x}) = r_e \rho(\bvec{x})$, where $\rho(\bvec{x})$ is the
complex electron density and $r_e$ the classical electron radius.  This object
scattering function may be decomposed into a Fourier representation of 3D spatial
frequencies \bvec{u}, with complex amplitudes
\begin{equation}
  \label{eq:FT}
  O(\bvec{u})= \mathcal{F}\{o(\bvec{x})\} \equiv \int o(\bvec{x}) \exp
  (2 \pi \, i \bvec{u} \cdot
  \bvec{x}) \, \mathrm{d}\bvec{x},
\end{equation}
in which spatial frequency can be thought of as a volume grating.  In
the case of coherent diffraction imaging a plane wave with wave-vector
$\bvec{k}_\mathrm{in}$ is incident on the object and the intensity of
the scattered field in the direction of the wave-vector
$\bvec{k}_\mathrm{out}$ is measured on a 2D pixellated detector
(e.g. a bare CCD) in the diffraction far field.  This detector is
typically centered on the forward direction, but in principle could be
oriented in any angle to the incident beam (see
Fig.~\ref{fig:geometry}).  For elastic scattering only the volume
gratings that satisfy Bragg's law will scatter, and the wave-vector
transfer $\bvec{q} = \bvec{k}_\mathrm{out} - \bvec{k}_\mathrm{in}$
will be equal to the grating spatial frequency; $\bvec{q} = \bvec{u}$.
Since the magnitudes $|\bvec{k}_\mathrm{out}|$ and
$|\bvec{k}_\mathrm{in}|$ are constant and equal to $1/\lambda$, these
spatial frequencies $\bvec{u}$ lie on the Ewald sphere of radius
$1/\lambda$
\cite{James:1962,Wolf:1969}, where $\lambda$ is the X-ray wavelength.
This construction is equivalent to the condition that to scatter light by an angle $2
\theta$ from the forward direction (the $z$ axis), the volume grating must be tilted by an
angle $\theta$ from perpendicular to the forward direction (Bragg's
law).  With the convention used here we have $|\bvec{q}| = q =
2/\lambda \sin \theta$.  The diffraction amplitudes in the direction
$\bvec{k}_\mathrm{out}$ are proportional to $O(\bvec{q})$, and in
diffraction imaging we measure the intensities, proportional to
$|O(\bvec{q})|^2$.  In particular, in the Born approximation (which
can be thought of in this context as single scattering), the number of
photons per second measured in a CCD pixel, with solid angle $\Omega$,
is given by
\begin{equation}
  \label{eq:intensity}
  I(\bvec{q}; \Omega) = I_0 \, \Omega \, P \, |O(\bvec{q})|^2,
\end{equation}
where $I_0$ is the flux (photons per second per unit area) of the
incident plane wave on the sample, and $P$ is the polarization factor;
$P = (1 + \cos^2 \psi)/2$ for unpolarized light, with $\psi = 2
\theta$ \cite{James:1962}.  

The complex scattering potential $o(\bvec{x})$ that we aim to recover from measurements of
$I(\bvec{q})$ is related to the complex refractive index $n (\bvec{x})$ of the object by
\cite[Sec.~13.1]{Born:2002} \cite{Kirz:1995}
\begin{equation}
  \label{eq:object}
    o(\bvec{x}) = r_e \rho (\bvec{x}) = \frac{\pi}{\lambda^2} \left ( 1 - n^2(\bvec{x}) \right
    ). 
\end{equation}
In the soft X-ray region, the complex refractive index is usually
written in terms of the optical constants as $n(\bvec{x})=1 -
\delta(\bvec{x}) - i \beta(\bvec{x})$. For optical constants much less
than unity, which is generally the case for soft X-rays,
Eqn.~\eqref{eq:object} can then be well approximated by
\begin{equation}
  \label{eq:object-approx}
  o(\bvec{x}) \approx \frac{2 \pi}{\lambda^2} \left ( \delta (\bvec{x}) + i \beta(\bvec{x})
  \right ) = \frac{2 \pi}{\lambda^2} \Delta n (\bvec{x}).
\end{equation}
The validity of Eqn.~\eqref{eq:intensity} under the Born approximation is that $D
|\Delta n (\bvec{x})| < 2 \pi \, \lambda \, C$, where $D$ is the thickness of the
object and $C \approx 0.2$ \cite{Natterer:2004}. 
 
\subsection{Experimental Requirements}
\label{sec:requirements}
The recovery of the 3D image $o(\bvec{x})$ from $O(\bvec{u})$ requires
the phases of $O(\bvec{u})$ to be recovered and combined with the
square root of the measured intensities.  Both phase retrieval and
image recovery place requirements on the 3D sampling of the
diffraction intensities.  Image recovery requires that the object be
adequately sampled in real space to resolve the finest desired feature
size over a given field of view.  The requirements of the phase
retrieval step are more demanding, in particular because our phase
retrieval algorithm has the additional task of overcoming gaps and
missing regions in the 3D sampled data, by performing the
interpolation tasks of recovering intensities that were blocked by a
beam-stop or that were missed due to a limited angular range of
measurements.  The 3D image recovery requires measuring the complex
amplitudes $O(\bvec{u})$ throughout a volume of reciprocal space.
Since a single diffraction pattern is limited to frequencies 
$\bvec{u}=\bvec{q}$ on the Ewald sphere, diffraction data
must be collected for various orientations of the sample.

In this work we perform phase retrieval and image recovery by full 3D
Fourier synthesis, which requires interpolating the measured
intensities from the Ewald sphere onto a uniform 3D Cartesian grid.
In reciprocal space the grid has a width of $N$ samples, spaced by
$\Delta u$, and is centered at the zero spatial frequency.  For $N$
even, the spatial frequencies along each grid axis run from
$-(N/2-1)\,\Delta u$ to $(N/2)\,\Delta u$.  In real space we
characterize the grid by a spacing $\Delta x$ and a field width $w = N
\Delta x$.  Since $\Delta x \, \Delta u = 1/N$ we have the
relationship $\Delta u = 1/w$, thus the largest spatial frequency
component along a grid axis is given by $u_{x,\mathrm{max}} =
N\,\Delta u/2 = 1/(2 \Delta x)$.

From Eqns.~(\ref{eq:FT}) and (\ref{eq:intensity}), the inverse Fourier
transform of the intensity diffraction pattern is proportional to the
autocorrelation function of the image that would be recovered when the
phases are known:
\begin{equation}
  \label{eq:autocorrelation}
  i(\bvec{x}) = \mathcal{F}^{-1}\{I(\bvec{q})\} \propto
  o(\bvec{x}) \otimes \cc{o}(\bvec{x}).
\end{equation}
Consider an object of finite extent with a maximum width $D$ along any
one axis of the real-space grid.  The autocorrelation image
$i(\bvec{x})$ in that direction has a maximum width of $2D$, and hence
the diffraction intensities are band-limited. That is, the smallest
grid spacing required to record all information present in the
diffraction intensities is $\Delta u = 1/(2D)$, the Nyquist critical
sampling frequency, and finer samples can be evaluated by a
sinc-series expansion of the measured samples \cite{Bracewell:1986}.
We define the sampling ratio $s$ per dimension with $w = s\,D$,
relative to Bragg sampling (band-limited sampling of intensities
occurs for $s=2$). The oversampling of data relative to the Bragg
sampling of $\Delta u = 1/D$ is what enables the phase retrieval
techniques to be employed. In practice we may measure data on a finer
grid than strictly required as a way to increase detector dynamic
range, although successful phase retrieval can also be achieved with
less than this factor of two in each dimension \cite{Miao:2003}.

The CCD must be placed so that it intersects large enough range of
scattering angles up to the desired spatial resolution.  Usually the
CCD is far enough away from the sample to be in the diffraction
far-field, in which the angularly-resolved diffraction pattern does
not vary with propagation distance. For an object of width $D$ the far
field exists beyond distances of $z_F = 2 D^2/\lambda$ from the object
\cite{Goodman:1996}. For a detector with pixels of width $p$ placed a
distance $z_D$ from the object, we have, for small scattering angles,
$\Delta q = p/(z_D \lambda)$.  That is, to sample a field width of $w
= s\,D$ the detector must be placed a distance of $z_D = s p D
/\lambda$.  This will be in the far-field if $z_D > z_F$, which can be
satisfied if $D < s p/2$, or the condition that the sample must be
smaller than the CCD pixel spacing when $s=2$. If the CCD is closer to
the sample than $z_F$ then the sample and diffraction planes are
related by a Fresnel, rather than a Fourier, transform, and the
reconstruction algorithms must be appropriately modified.

Experimental requirements are placed on the transverse and
longitudinal coherence of the incident beam.  The transverse spatial
coherence length of the incident beam must be at least as large as the
entire field width $w = s\,D$ \cite{Spence:2004a}. The effect of
partial coherence may be modeled as an incoherent source located some
distance from the sample, whereby the diffraction intensity pattern is
convolved with a demagnified intensity image of the source.  In real
space this convolution modulates the autocorrelation of the object
with an envelope function which is proportional to the modulus of the
transform of the source function.  By the Van Cittert-Zernike theorem
\cite{Born:2002}, this envelope function is the mutual coherence of
the source.  The measured diffraction intensity is also convolved with
the pixel response function of the detector, which modulates the
autocorrelation image with an envelope proportional to the MTF of the
detector. The spectral bandwidth $\Delta \lambda/\lambda$ of the
incident light should be narrower than $2/N$ since we require $\Delta
q/q = \Delta
\lambda/(2\lambda)$ so that the range of angles diffracted from a
single spatial frequency by a range of wavelengths spreads by no more
than half a detector pixel. This is equivalent to a minimum required longitudinal
coherence length of $w q_\mathrm{max} \lambda = 2 w \sin \theta$, which will be the
maximum path-length for light scattering by $2\theta$ to the edge of
the detector from points
spaced transversely by $w$, or by the same angle from points spaced
longitudinally by $w / \tan 2\theta$.

In our experiments we rotate the sample about an axis perpendicular to
the incident beam direction to build up the 3D dataset.  At the
highest spatial frequencies recorded, an angular increment of $\Delta
\phi$ leads to a spacing between the Ewald sphere surfaces of $\Delta
q = q_\mathrm{max} \, \Delta \phi$.  That is, the Crowther resolution
\cite{Crowther:1970} matches the critical sampling of the diffraction
intensities ($s=2$) when 
\begin{equation}
  \label{eq:Crowther}
  \Delta \phi = \Delta q/q_\mathrm{max} = \Delta x/D.
\end{equation}
Note that this angular increment leads to a higher than necessary
sampling at the lower spatial frequencies.  For the examples in this
paper we collected diffraction data with angular increments that are
2--4 times larger than given by Eq.~(\ref{eq:Crowther}). In the
process of phase retrieval we additionally recover both the amplitudes
and phases of the missing data between the Ewald surfaces, including
those in a large gap resulting from a limited range (usually $\pm
70\degree$) of rotation angles, data blocked by a beam-stop, and the
missing ``cone'' of data resulting from rotating the sample about a
single axis.  This amplitude and phase retrieval of the missing data
is essentially a super-resolution technique and is achieved with the
same real-space image constraints that we use for phase retrieval
\cite{Salerno:1998}.  Recovery of unmeasured data behind a beamstop
has been demonstrated previously in 2D coherent X-ray diffraction
imaging using this technique \cite{Marchesini:2003, Nishino:2003}, and
data missing due to a limited number of views have been recovered in
the context of computed tomography using iterative algorithms
\cite{Sato:1981} similar to those we use (described in 
Sec.~\ref{sec:phase-retrieval}).
Depending on how much amplitude information is missing, there will be
a null space of images which are not constrained by the real or
reciprocal space constraints \cite{Bertero:1982, Salerno:1998} and
which may need to be regularized in the reconstruction \cite{Shapiro:2005}.

\subsection{Interpolation of the Diffraction Intensities}
\label{sec:interpolation}
We interpolate the diffraction intensities measured on the CCD
detector onto a regular Cartesian grid.  The location of the CCD pixel
indexed by integers $(i, j)$ is given by the vector $\bvec{p}_{i,j} =
p(i \,
\bvec{\hat{i}} + j \,
\bvec{\hat{j}})$, relative to the pixel intersecting the direct beam, as shown in Fig.~\ref{fig:geometry}.
We have then, with $\bvec{k}_\mathrm{in} = (1/\lambda) \bvec{\hat{k}}$, 
\begin{equation}
  \label{eq:CCD_q_ij}
  \bvec{q}_{i,j} = \bvec{k}_\mathrm{out}-\bvec{k}_\mathrm{in}  = 
  \frac{1}{\lambda} \, 
  \left ( \frac{\bvec{p}_{i,j} + z_D \bvec{\hat{k}}}{\sqrt{p_{i,j}^2+z_D^2}} -
    \bvec{\hat{k}} \right ),
\end{equation}
where $z_D$ is the distance from the sample to the detector.  Hence, for
example,
\begin{align}
  \label{eq:CCD_q_x}
  q^x_{i,j} &= \bvec{q}_{i,j} \cdot \bvec{\hat{i}} = 
  \frac{1}{\lambda} \, \frac{p \, i}{\sqrt{p^2
      (i^2+j^2)+z_D^2}}, \\
  q^z_{i,j} &= \bvec{q}_{i,j} \cdot \bvec{\hat{k}} = 
  \frac{1}{\lambda} \, \left ( \frac{z_D}{\sqrt{p^2
      (i^2+j^2)+z_D^2}} - 1 \right ).
\end{align}
In practice each diffraction pattern in our data set has a different
rotation in the 3D Fourier space of the object, and the coordinate of
each pixel in this space is given by
\begin{equation}
  \label{eq:u}
  \bvec{u}_{i,j,\phi} = R_\phi \, \bvec{q}_{i,j},
\end{equation}
where $R_\phi$ is the 3D rotation matrix derived from the known
object orientation.  The coordinates $\bvec{u}_{i,j,\phi}$ are
then mapped onto a uniform Cartesian grid by nearest-neighbor
sampling.  Where more than one pixel from the set of intensity
measurements contribute to a given voxel, the pixel values are
averaged to determine the appropriate intensity value at that point.

We note that there are alternatives to the nearest-neighbor
interpolation onto a regular grid that we use in this
work.  The interpolation could be avoided by solving the inverse
transform by inverse methods, such as performed in the computer program
\textsc{back} \cite[Sec.~A5.3]{Szoke:1997} which utilizes a
constrained conjugate gradient solver and which is used in the
computer program \textsc{speden} \cite{Hau-Riege:2004}
(\textsc{speden} is a program to perform 3D phase retrieval, designed
specifically to optimally include prior data and avoid instabilities
caused by interpolation.)  Alternatively, it should be possible to
employ fast algorithms to compute discrete Fourier transforms of
non-equispaced data (NDFTs)
\cite{Potts:2001}.  In the non-diffracting limit (i.e.  computed tomography, or
CT) the reconstruction method of filtered back-projection can be shown
to be a computationally efficient method that is equivalent to Fourier
synthesis via a polar-to-Cartesian interpolation \cite{Choi:1998,
  Natterer:2001}.  A corresponding algorithm, called filtered
back-propagation \cite{Devaney:1982}, has been developed for the
diffracting case where the diffraction amplitude and phase are
measured, which again can be shown to be equivalent to Fourier
synthesis after interpolation \cite{Pan:1983}.

\subsection{Phase Retrieval}
\label{sec:phase-retrieval}
Our phase retrieval method follows from previous work on 2D diffraction imaging
\cite{He:2003a,He:2003,Marchesini:2003a,Marchesini:2003}.  In particular, we have extended the Shrinkwrap
algorithm \cite{Marchesini:2003} to operate on 3D arrays. This
algorithm is based on an iterative transform algorithm (ITA), which
cycles between real and reciprocal space, respectively enforcing the
constraints of known object support or known diffraction
modulus. Usually an ITA requires knowledge about the shape of the
object to set the support constraint.  This support is usually larger
than the actual boundary of the object; what is termed a loose
support.  For general complex-valued objects, where a positivity
constraint can not be applied, the ITA gives higher-quality
reconstructions when the support constraint more closely and tightly
matches the object's boundary \cite{Fienup:1987}.  The reason for this
is explained in Sec.~\ref{sec:2D}.  The Shrinkwrap algorithm
periodically updates the estimate of the support based on the current
object estimate.  The updated support is chosen by low-pass filtering
the current estimate and setting the support to be the region for
which the intensity is above a certain threshold (usually a prescribed
fraction of the maximum image intensity).  The method can be started
from an estimate of a very loose support, from a threshold of the
object's autocorrelation function, or even the entire array.  A method
which exists for finding an estimate of the object support from the
autocorrelation function's support could also be used
\cite{Fienup:2005}.  While the Shrinkwrap method can be used with any
ITA, such as the Hybrid Input-Output (HIO)
\cite{Fienup:1982} or Difference Map \cite{Elser:2003} algorithms, we
used the HIO and Relaxed Averaged Alternating Reflections (RAAR)
\cite{Luke:2005} algorithms for this work.

Many of the phase retrieval ITAs can be written as fixed point
iterative equations, which can be written generally in the form
$g_{n+1} = \mathcal{T} g_n$, for a generic operator $\mathcal{T}$.
The RAAR algorithm can be represented by the fixed point iterative
equation of the complex-valued real-space image iterate $g$
\cite{Luke:2005}:
\begin{equation}
  \begin{split}
  \label{eq:RAAR}
  g_{n+1} & = \left[ \frac{1}{2} \beta (R_S \, R_M + I) +(1-\beta) P_M \right] g_n \\
  & = \left [ 2 \beta \, P_S \, P_M  + (1-2 \beta) P_M + \beta (P_S
    -I)\right ] g_n,
  \end{split}
\end{equation}
where the operator $R = 2P - I$ is the reflector corresponding to the
projector $P$, $I$ is the identity operator, and $\beta$ is a feedback
parameter, which we usually set to $\beta = 0.9$.  The two operators
$P_M$ and $P_S$ are the projections onto the modulus constraint and
support constraint, respectively.  We apply the modulus according to
\begin{widetext}
\begin{equation}
  \label{eq:modulus}
  P_M \, g = \mathcal{F}^{-1}
  \begin{cases}
    \frac{G(\bvec{u})}{|G(\bvec{u})|+\epsilon}
      \left ( \sqrt{I(\bvec{u})} + \sigma_{\sqrt{I}}(\bvec{u}) \right),
    & \text{if $G(\bvec{u}) > \sqrt{I(\bvec{u})}+ \sigma_{\sqrt{I}}(\bvec{u})$,} \\
    \frac{G(\bvec{u})}{|G(\bvec{u})|+\epsilon}
      \left ( \sqrt{I(\bvec{u})} - \sigma_{\sqrt{I}}(\bvec{u}) \right),
    & \text{if $G(\bvec{u}) < \sqrt{I(\bvec{u})}- \sigma_{\sqrt{I}}(\bvec{u})$,} \\
    G(\bvec{u}), & \text{otherwise, or } u \not \in M
  \end{cases}
\end{equation}
\end{widetext}
where $\sigma_{\sqrt{I}}$ is the estimated variance of the measured
diffraction amplitudes, $G(\bvec{u}) = \mathcal{F} \{ g(\bvec{x}) \}$,
and $\epsilon$ is a small number.  $M$ is the set of $\bvec{u}$ where
$I(\bvec{u})$ has been measured.  For example, $\bvec{u} \not\in M$ in
the missing sector of data present when diffraction is recorded over a
limited range of angles.  The operator $P_M$ of
Eqn.~\eqref{eq:modulus} retains the phase of the complex Fourier
amplitude $G(\bvec{u})$ and projects its modulus $|G(\bvec{u})|$ to
the nearest measured diffraction amplitude, in the interval
$[\sqrt{I(\bvec{u})} - \sigma_{\sqrt{I}}(\bvec{u}), \sqrt{I(\bvec{u})}
+
\sigma_{\sqrt{I}}(\bvec{u})]$ (or does nothing if the modulus already lies within that
range or if $I(\bvec{u})$ has not been measured).  Given the support
$S$ from Shrinkwrap, we apply the support constraint when retrieving
the phase of a complex image using
\begin{equation}
  \label{eq:support}
  P_S \, g = 
  \begin{cases}
    g(\bvec{x}) & \text{if $\bvec{x} \in S$} \\
    0 & \text{otherwise.} 
  \end{cases}
\end{equation}
We also perform phase retrieval where we impose real and positive constraints on the image
amplitudes, where we replace $P_S$ with
\begin{equation}
  \label{eq:support-positive}
  P_{S+} \, g = 
  \begin{cases}
    \Re \{g(\bvec{x})\} & \text{if $\bvec{x} \in S$ and $\Re \{g(\bvec{x})\} > 0$} \\
    0 & \text{otherwise.}
  \end{cases}
\end{equation}

The HIO algorithm can only be written in terms of a fixed point
iterative equation when applying the support constraint $P_S$, but not
when applying positivity constraints \cite{Luke:2005}.  In general the
HIO algorithm is given by
\begin{equation}
  \label{eq:HIO}
  g_{n+1} = 
  \begin{cases}
    P_M \, g_n, & \text{if $\bvec{x} \in S'$} \\
    (I - \beta \, P_M) g_n, & \text{otherwise,}
  \end{cases}
\end{equation}
where $S'$ is the set of elements where $P_M \, g_n$ satisfies the
support and (if desired) the reality and positivity constraints.  As
with the RAAR algorithm we
use a value of the feedback parameter $\beta = 0.9$. 

Regardless of algorithm, we monitor the reconstruction with the real-space image error
\begin{equation}
  \label{eq:error}
  E_S^2 \equiv \frac{\sum \left| g_n - P_S \, g_n \right|^2}{\sum \left| P_S \, g_n \right|^2}
  = \frac{\sum_{\bvec{x} \not{\in} S} \left| g_n(\bvec{x}) \right|^2}{\sum_{\bvec{x}
      \in S} \left| g_n(\bvec{x}) \right|^2}.
\end{equation}
This metric is a measure of the total power in the image that remains
outside the support, and is zero for the case of perfectly satisfying
the real-space constraints.  We define, in a completely analogous way
to Eqn.~\eqref{eq:error}, the error $E_M$ corresponding to the modulus
constraint defined by $P_M$:
\begin{equation}
  \label{eq:modulus-error}
  E_M^2 \equiv \frac{\sum \left| g_n - P_M \, g_n \right|^2}{\sum \left| P_M \, g_n \right|^2} = 
  \frac{\sum \left| |G_n| - \sqrt{I}\right|^2}{\sum I},
\end{equation}
where the equality follows from Parseval's theorem and is true only for $\sigma_{\sqrt{I}}
= 0$.
The error metrics $E_S$ and $E_M$ are the normalized distances between the
current iterate $g_n$ and the support or modulus constraint set, respectively.
The reconstructed image from a reconstruction run (from a particular set of starting
phases) is given by 
\begin{equation}
  \label{eq:image}
  \gamma_M = P_M \ g_n,
\end{equation}
for the final iterate $g_n$ of both the RAAR and HIO algorithms.

The Shrinkwrap algorithm has been used previously to reconstruct 2D
images of thin objects at a resolution of about 20\nm
\cite{Marchesini:2003}.  We have found in subsequent studies that the
step of updating the support would sometimes shrink the support to a
shape smaller than the actual boundary of the object.  To counter this
effect we have improved the Shrinkwrap algorithm to prevent it from
over-shrinking the support.  Depending on the parameters of the
low-pass filter and the threshold level, the support may start to cut
off the extremities of the object.  At this point the support
constraint error $E_S^2$ increases rapidly and the reconstructions
rapidly degrade with further iteration.  This error is thus a good
indicator of when the halt the support refinement.  We simply monitor
the error metric and when it increases above a set point we choose the
support saved from 10 iterations prior.  This then becomes our best
estimate of the support and is used as a fixed support in combination
with the RAAR algorithm for many more (typically 100 to 1000)
iterations.  We further decrease the uncertainty of the retrieved
phases by averaging the retrieved complex images from independent and
random starting diffraction phases using the Shrinkwrap-derived
support constraint \cite{Shapiro:2005} as described in
Eqn.~(\ref{eq:image}) of Sec.~\ref{sec:resolution}.  If the phase at a
particular spatial frequency is randomly recovered from trial to
trial, the average modulus will average to zero, and hence be filtered
out of the recovered image.

The 2D reconstructions shown in this paper were reconstructed using
the RAAR algorithm (Eqn.~\ref{eq:RAAR}) and the 3D were performed
using a combination of HIO (Eqn.~\ref{eq:HIO}) and RAAR.  A typical
reconstruction process proceeds as follows.  First we define the
initial object support mask by applying a 2\% intensity threshold to
the object autocorrelation, obtained by Fourier transforming the
measured diffraction pattern. The support constraint, defined by the
current object mask, is applied to the solution in real space once per
iteration.  We typically use a feedback parameter of $\beta=0.9$ in
the RAAR or HIO algorithms.  The object support $S$ is recomputed
every 30 iterations by convolving the absolute value of the current
reconstruction $\gamma_M$ with a Gaussian of FWHM of initially three
pixels in all dimensions and applying a threshold to the resultant
image at 15\% of the maximum value.  As the iterations progress we
reduce the width of the Gaussian blurring function from three pixels
to one pixel, following the prescription $w_S = 1+2 \exp
(-n^2/n_w^2)$, with $n_w$ regulating the speed at which $w_S$
decreases with iteration number $n$.  The reduction in the blurring
width enables the support to better conform to the solution as the
quality of the reconstruction increases.  We perform this Shrinkwrap
support determination without applying any real-space positivity or
reality constraint on the image amplitudes (that is, we use the
constraint $P_S$ in the RAAR algorithm, or $S'=S$ in the HIO
algorithm).  The final support is usually obtained after 300 to 600
iterations, with a stopping criterion that the support constraint
error $E_S^2$ does not exceed 0.2.  Once the support is determined we
carry out many iterations of the RAAR algorithm, starting from random
phases, using a feedback parameter of $\beta=0.9$.  In some cases,
additional real-space constraints, such as positivity or reality of
the image amplitudes, are also applied.

As shown in Eqn.~(\ref{eq:modulus}), in diffraction space the
amplitudes of the object guess are matched in magnitude to the
measured diffraction pattern amplitude over those parts of 3D
diffraction space where the measured intensity is defined.  Those
parts of 3D diffraction space where there is no measured data are
allowed to float and are not constrained. This includes the regions
between the measured Ewald spheres, the missing wedge of data from the
finite range of rotation angles, the central beamstop region, and
those parts of the diffraction pattern where the measured intensity is
sufficiently low to be regarded as noise.  An additional, optional
Fourier space constraint is to set those pixels beyond the radius of
the spatial frequencies measured by the CCD chip to zero.  This
asserts lack of knowledge of spatial frequencies higher than those
measured by the CCD camera, and effectively provides a pupil function
for the imaging system in three-dimensional space.

Providing an initial guess for the 3D object support is not typically
necessary but speeds the reconstruction process and helps break
inversion symmetry present in the object autocorrelation.  An initial
3D support estimate can be obtained from the diffraction data by first
performing Shrinkwrap phase retrieval on a 2D central section, as
described in Sec~\ref{sec:2D-projection}.  We then extrude the 2D
support mask that was generated into 3D to provide an initial 3D
support estimate. If several 2D reconstructions are available from a
range of views, the intersection of these support functions in 3D can
be used to provide a more detailed initial support
estimate. Experience has shown that even a low-resolution or
comparatively poor support estimate is sufficient to almost
immediately break any inversion symmetry in the reconstruction and
hasten convergence of the 3D solution. Performing such a 2D
reconstruction is a common (although not strictly necessary) step in
assessing data quality prior to performing 3D reconstruction.

\section{Methods}
\subsection{Sample Preparation}
\label{sec:samples}
A goal of this study was to be able to unambiguously compare
reconstructed X-ray images of a three-dimensional object with images
obtained by another high-resolution method, such as a scanning
electron microscope (SEM).  To accomplish this we fabricated a test
object that consists of a silicon nitride membrane with a
three-dimensional pyramid shape that is decorated with 50-nm-diameter
colloidal gold spheres, similar to that previously described
\cite{Marchesini:2003a}.  The
object is three-dimensional and has a comparable width, height, and depth,
measuring 2.5\micron $\times$ 2.5\micron $\times$ 1.8\micron.

The pyramid-shaped membrane was fabricated by lithography using
methods similar to those to make silicon nitride windows and silicon
nitride atomic-force microscope (AFM) tips.  The starting material was
a double-side polished 200\micron thick wafer of silicon crystal with
the crystal 100 axis oriented normal to the surface.  Pits with an
inverted pyramid shape were etched into one side of the wafer by
anisotropic etching through a pattern of 2.5\micron-width square
holes, lithographically printed and developed in photo-resist.  The
anisotropic etch leaves the 111 crystal planes exposed, so that the
surface normal of any one of the four faces of the pyramid makes an
angle of 54.7\degree to the window normal and the ratio of the depth
of the pit to its base width is $1/\sqrt{2}$.  After removing the
photoresist a low-stress silicon nitride film of 100 nm thickness was
grown on the surface by chemical vapor deposition.  Window openings
were then etched from the other side of the wafer after first resist
coating and patterning that side, making sure to align to marks etched
in the front surface.  The etch from the back removes silicon, but
leaves a free-standing membrane of silicon nitride, which in this case
had one pyramid-shaped indentation per window.  The windows were made
with a slotted shape of about 2 mm width by 50\micron high.  With the
200\micron thickness of the silicon frame and the pyramid positioned
in the center of the window, this allows a line of sight through the
window at a maximum rotation angle (about an axis in the plane of the
window, parallel to the short window dimension) of 78\degree.

\begin{figure*}[htbp]
  \centering
  \includegraphics[width=0.8\textwidth]{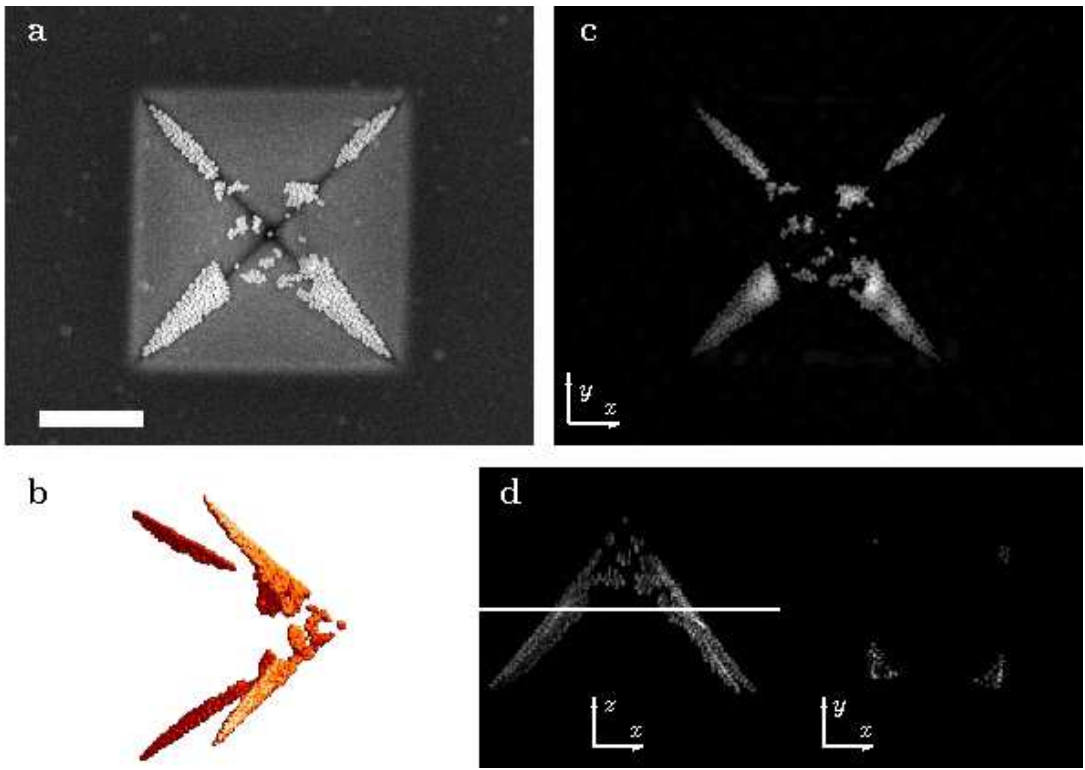}
  \caption[SEM image of the pyramid test object]{(Color online) (a) SEM image of the
    pyramid test object, consisting of 50-nm diameter gold spheres
    lining the inside of a pyramid-shaped indentation in a 100-nm
    thick silicon nitride membrane.  The membrane extends over a
    window of size $50\micron \times 1.7\mm$, the pyramid base width
    is 2.5\micron, and height is 1.8\micron.  (b) An
    iso-surface rendering of the reconstructed 3D image.  (c) Infinite depth of field X-ray
    projection image from a central section of the 3D diffraction
    dataset, reconstructed using the Shrinkwrap algorithm. (d) A
    maximum-value projection of the 3D reconstructed image (left) with
    a vertical white line indicating the location of a tomographic
    slice (right). The scale-bar
    length is 1\micron and applies to all images.}
  \label{fig:SEM}
\end{figure*}

The gold-sphere test object was made by dragging a small drop of
solution of gold balls in water, suspended from a micro-pipette,
across the silicon nitride window so that it intersected with the
pyramid indentation.  Best success was achieved with a slightly
hydrophilic silicon nitride surface, which could be obtained by
cleaning the surface in an oxygen plasma.
As the drop was moved over and away from the indentation, a smaller
drop broke away from the main drop and was captured in the pyramid.
This drop quickly evaporated and left the gold balls in a
characteristic pattern where the gold tended to fill in the edges of
the pyramid.  The main drop was completely dragged away from the
window, so the only gold balls on the window were those in the
pyramid.  A plan-view SEM image (membrane and wafer perpendicular to
the electron beam) of the object is shown in Fig.~\ref{fig:SEM}.  The
SEM is however only sensitive to the surface of the object---the
electrons do not penetrate the gold spheres nor the membrane.  The
depth of focus of the SEM was larger than the thickness of the object,
and from the plan view we can determine the lateral coordinates of the
topmost balls and infer the third coordinate from the known geometry
of the pyramid.

The silicon nitride window was glued to a pin so that the pyramid was close (within about
20\micron) of the rotation axis of the pin.  The pin was mounted in a collar that was
attached to a JOEL electron microscope sample holder.  This assembly was mounted into the
modified goniometer holder of the diffraction apparatus \cite{Beetz:2005}.
  
\subsection{Data Acquisition}
\label{sec:acquisition}
Experiments were carried out at an undulator source at the Advanced
Light Source (ALS) using the Stony Brook University diffraction
apparatus \cite{Beetz:2005}.  Pertinent to this experiment, 750 eV
(1.65 nm wavelength) X-rays were selected from the undulator radiation
by a zone-plate monochromator with a spectral resolution of
$\lambda/\Delta
\lambda = 1000$.  The 5-$\mu$m-diameter
monochromator exit pinhole also selects a transversely spatial
coherent patch of the beam.  The sample was located 20 mm from this
pinhole.  A direct-detection bare CCD detector, with 20\micron pixel
spacing, $1340 \times 1300$ pixels, was located 142 mm behind the
sample.  We selected sub-arrays of $1200 \times 1200$ elements,
centered on the location of the zero spatial frequency (direct beam).
At these CCD and wavelength settings we have a real-space sampling
interval in $x$ and $y$ of $\Delta x = 9.8\nm$ (in the small-angle
approximation) and a field width of $w = N \Delta x = 11.7\micron$.
With these settings the 2.5\micron-wide pyramid object satisfies the
far-field and sampling conditions discussed in
Sec.~\ref{sec:requirements}.  The diffraction from the pyramid object
is more than $4 \times$ oversampled in each dimension ($s=4.6$).

The frame of the slotted window in which the pyramid is formed 
blocks most of the high-angle scatter from the pinhole
that would otherwise illuminate the CCD.  This scatter reveals a
projection shadow image of the slotted window, useful for aligning the
pyramid to the beam.  The diffraction pattern of the pyramid measured
by the CCD is shielded from this remaining pinhole scatter with a
small aperture placed 6\mm upstream of the sample (a distance at
which the sample can be rotated without interference).  A beam-stop
blocks the direct undiffracted beam from impinging on the CCD.  More
details are given by Beetz et al. \cite{Beetz:2005}.

\begin{figure}[htbp]
  \centering
  \includegraphics[width=0.4\textwidth]{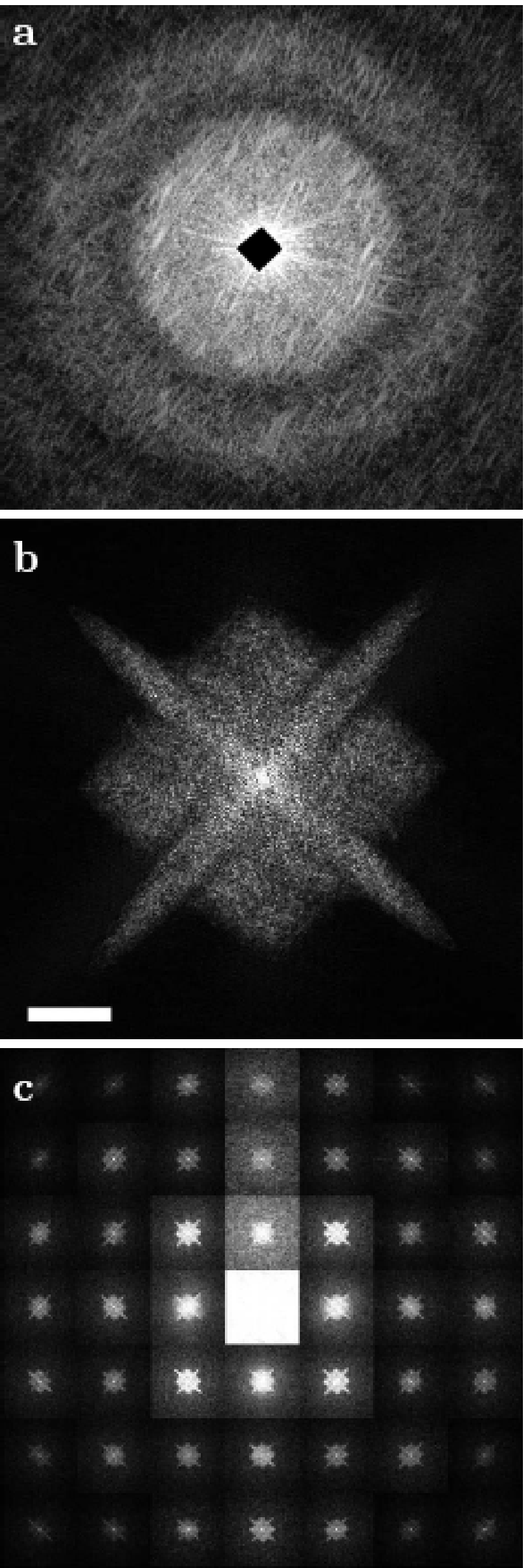}
  \caption[Autocorrelation images]{(a) The diffraction pattern for the
    $\phi = 0\degree$ orientation of the pyramid. (b) Autocorrelation image $i(\bvec{x})$, formed by
    Fourier transforming the diffraction intensity pattern of
    (a) after filtering, displayed with a linear
    greyscale (white highest intensity). 
    Scale bar is 1\micron. (c) Autocorrelation spectrogram of the same
    single-view diffraction pattern of the pyramid, displayed with a
    logarithmic greyscale.}
  \label{fig:spectrograms}
\end{figure}

Diffraction patterns were collected with the sample oriented at
rotation angles of $-57\degree$ to $+72\degree$, at 1\degree intervals
(compared with 0.27\degree angular increments required for full
sampling according to Eqn.~\eqref{eq:Crowther}).  The shadow of the
sample support frame limited useful data to $-57\degree$ to
$+66\degree$.  We additionally
collected data at 0.5\degree increments for a range of 19\degree
centered at an object orientation of $\phi = -26\degree$ from the head on
($\phi = 0\degree$) orientation.  
To keep the sample
centered in the 5\micron beam, the position of the sample was
calibrated by performing a two-dimensional raster scan of the rotation
and $y$ goniometer motors.  The total scattered counts (not
including those blocked by the beam-stop) were collected for each motor
position and the optimum $y$ position (a translation motion
perpendicular to the rotation axis) was then computed for each
rotation angle, and these were fit to a smooth curve as a function of
rotation angle.  To collect the 3D dataset, at each rotation angle we took several exposures to
accommodate the large dynamic range of $10^5$ of the diffraction pattern, and to
reduce the area occluded by the beam-stop (by setting the beam-stop to
different positions).  After subtracting dark noise, pixel data that
were not saturated and not masked by the beam-stop were summed over
these exposures, and then normalized by the accumulated incident flux
corresponding to that sum.  A typical diffraction pattern is shown in
Fig.~\ref{fig:spectrograms} (a), which was composed of 10 or more individual exposures
of 0.1\s, 1\s, 10\s, and
60\s duration, for a cumulative exposure of 73\s.  The diffraction pattern intensities are displayed on a
logarithmic greyscale in Fig.~\ref{fig:spectrograms}.  At the highest
angles of the pattern (highest resolution, at 0.07\invnm, along the diagonal) the mean photon count is
1.9 photons per pixel for this sample orientation.  The maximum normalized photon count, which occurs in a pixel near
the center of the pattern is 109,000 photons.  The estimated incident
flux was $8 \times 10^9$  photons/s/$\mu \text{m}^{2}$ (per 400\mA of
storage-ring current), and the
normalized incident fluence for the accumulated sum of
Fig.~\ref{fig:spectrograms} (a) was $3 \times 10^{11}$ photons/$\mu
\text{m}^{2}$.  The total normalized scattered counts at the
CCD over the accumulated exposure time for the pattern in
Fig.~\ref{fig:spectrograms} (a) was $1.6 \times 10^8$ photons (equal
to the total counts that would be recorded if the detector had
infinite dynamic range and did not saturate).

\begin{figure}[htbp]
  \centering
  \includegraphics[width=0.4\textwidth]{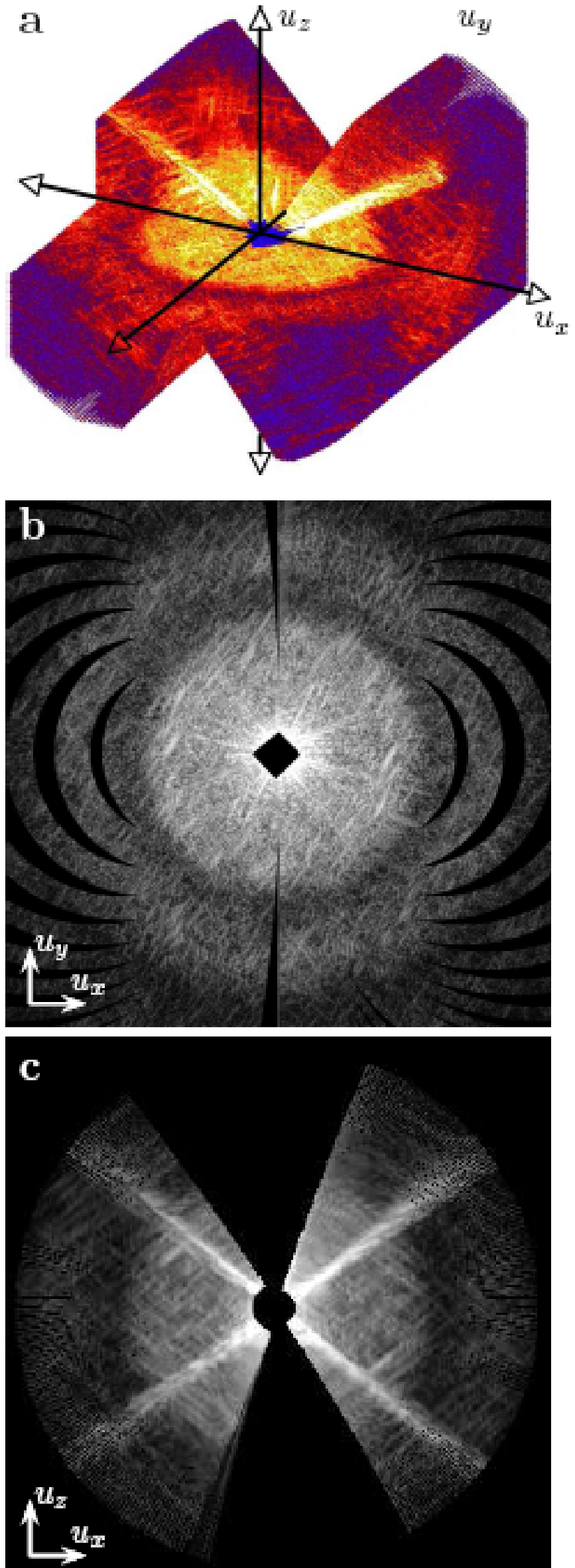}
  \caption[Diffraction pattern]{(Color online) (a) A rendering of the
    entire 3D diffraction dataset. (b) A central slice of the dataset
    in the a plane rotated by $-24\degree$ about the $y$ axis from the
    $u_x$-$u_y$ plane. (c) A central slice of the dataset in the $u_x$-$u_z$
    plane.  All patterns are displayed on a logarithmic greyscale (white highest intensity). The
    half width of each pattern is $u_{x,\mathrm{max}} = 0.048\invnm$.}
  \label{fig:diffraction}
\end{figure}

Views of the diffraction data cube are shown in
Figs.~\ref{fig:diffraction} (a)--(c) and discussed in
Sec.~\ref{sec:3D-images}.  This cube was assembled from the
123 diffraction patterns at 1\degree sample orientation increments,
and 32 patterns at half-degree intervals, by
interpolating onto $\bvec{u}_{i,j,\phi}$. The total integrated exposure time for the complete
dataset was 3.2 hours, with a total incident fluence of $5 \times 10^{13}$ photons/$\mu
\text{m}^{2}$.

\subsection{Data Diagnostics}
\label{sec:spectrogram}
As seen in Eqn.~(\ref{eq:autocorrelation}) the autocorrelation of the
object can be determined from a single inverse Fourier transform of the
measured data without having to recover the diffraction phases.  
We find that computing the autocorrelation image from the measured
diffraction data is a useful diagnostic to determine if the
measurement fulfills the sampling requirements, to help identify the
object, and to assess the measurement quality.  The inverse Fourier
transform of the diffraction pattern shown in
Fig.~\ref{fig:spectrograms}~(a) is given in
Fig.~\ref{fig:spectrograms}~(b). The displayed autocorrelation image has been cropped by
half in width and height from the inverse Fourier transform of the diffraction pattern,
since with the linear greyscale displayed the rest of the field was black. This autocorrelation image has a
well-defined support which is confined within the image field, showing
that we are indeed oversampling the diffraction intensities.

The Fourier relationship of Eqn.~(\ref{eq:autocorrelation}) offers a
further method to diagnose the diffraction pattern $I(\bvec{q})$ as a
function of the spatial frequency $\bvec{q}$ across the pattern.  A
property of the Fourier transform of a function, often discussed in
the context of holography, is that a sub-region of the transform (or
hologram) can be inverse Fourier transformed to give a
spatial-filtered image of the original function.  The field of that
image is the full field of the original function.  The filtered image
may differ from sub-region to sub-region, since each sub-region
contains information pertaining to particular spatial frequencies of
the original object function.  Nevertheless, for non-periodic object
functions, these images should be consistent and not vary too
dramatically between neighboring regions.  Large inconsistencies in
images formed in neighboring regions point to inconsistencies in the
measured diffraction data.  This gives a convenient way to
qualitatively check the diffraction intensities (and also
reconstructed phases) across the diffraction pattern, by forming an
array of sub-images, each corresponding to a particular sub-region.
We term this array a ``spectrogram'', since it displays the image
information as a function of spatial frequency, much like the
time-frequency spectrograms used in audio analysis.  We apply the
spectrogram analysis both to the autocorrelation image, and to
reconstructed images to determine the consistency of the data or
reconstructed phases.

An example of an autocorrelation spectrogram is shown in
Fig.~\ref{fig:spectrograms}~(c), where each image is formed by
applying a 2D Gaussian window function to the diffraction pattern,
centered at a location indicated by the image position in the array.
One effect that is immediately noticeable to the eye is that, at the
higher resolution positions, the images vary according to their
azimuthal position in the spectrograph.  In particular features that
are tangential to circles of constant resolution have greater contrast
than features that vary in a radial direction.  The effect gives the
appearance that the spectrograph is rendered onto the surface a
sphere, but is in fact a signature of limited longitudinal coherence
of the incident beam.  For a given $\bvec{q}$, represented in the
spectrograph by the position of the image, pairs of points of the
object that are perpendicularly bisected by the vector $\bvec{q}$ will
interfere with zero path difference.  These points will interfere with
the same phase for all wavelengths (assuming no dispersion of the
scattering factors). The path difference at $\bvec{q}$ of other pairs
of points along this line perpendicular to $\bvec{q}$ depend only in
second order on their mean displacement from the direction of
$\bvec{q}$.  The path differences of rays scattering from pairs of
points separated in the radial direction, however, vary to first order
on their radial separation.  Therefore, a limited longitudinal
coherence, which limits the path difference over which interference
occurs, will reduce interference for points separated by a direction
parallel to the $\bvec{q}$ vector by a much greater extent than for
the perpendicular direction.  The spectrograph gives a good visual
determination of the degree of longitudinal coherence, and we see from
Fig.~\ref{fig:spectrograms}~(c) that the longitudinal coherence is
adequate for diffraction out to the center edge of the pattern, but
not quite adequate for the corner of the pattern.  By comparison to
spectrographs of simulated diffraction patterns, we estimate that the
relative bandwidth of the incident beam in this case is approximately
$\Delta \lambda/\lambda = 1/600$.

It is also clear from Fig.~\ref{fig:spectrograms}~(c) that the data
are inconsistent at the low spatial frequencies, since at those
$\bvec{q}$ positions there is noticeable intensity outside the support
of the pyramid object autocorrelation.  This is due to the fact that
low-frequency data are missing due to the beamstop, and also to a
lesser degree due to scattering from the sample substrate or the
coherence-defining pinhole.  The data are visibly noisier in windowed
regions located in the first three rows of the fourth column of
Fig.~\ref{fig:spectrograms}~(c), due to the stalk that holds the
beamstop and which was moved over several positions in this region for
the cumulative exposure.  The noise and inconsistency can cause the
Shrinkwrap algorithm to fail (in which it keeps shrinking beyond the
object's boundary), especially when applied without an image-space
constraint such as positivity.  We find the Shrinkwrap algorithm
consistently converges to a stable object support when we apply a
high-pass filter to the diffraction intensities prior to
interpolation.  This filter has the form
\begin{equation}
  \label{eq:filter}
  f(q) = 
  \begin{cases}
    (q/2 a)^4 \exp (2-q^2 / 2 a^2), & \text{if $q <
      2a$,} \\
    1, & \text{otherwise,}
  \end{cases}
\end{equation}
where $q = |\bvec{q}|$ and the filter radius $a$ is 100 pixels, or less than 10\% of the
array width.  The image thus formed will be a high-pass filtered
image, equivalent to the coherent image formed by a lens with a
central obscuration.  The filter was applied to the data of Fig.~\ref{fig:spectrograms}~(a), prior to
Fourier transformation, to diminish the effects of the beamstop.
This filter also regularizes the inverse
transform, which is ill-posed in part due to the missing data behind the
beamstop \cite{Bertero:1982, Salerno:1998}, by simply setting the
unknown intensities to be zero.  The effect of this high-pass filter
is to convolve the coherent image with the Fourier transform of the
filter. This causes a ringing of the image, which gives rise to
negative amplitudes in the image, and a slightly larger image support.
We also zero the diffraction intensities of the bright cross streaks
seen in the $x$-$z$ central section, to suppress artifacts that they may
cause.

\subsection{Computational Implementation}
\label{sec:computation}
The two key computational challenges in implementing high-resolution
3D phase retrieval at the time or writing are performing the numerous
3D FFTs required in a reasonable period of time and managing the
memory requirements of the large 3D data arrays.

Memory requirements are dictated by the size of the data sets acquired
and by the phase retrieval algorithms used.  For the iterative
transform phase retrieval methods described in
Sec.~\ref{sec:phase-retrieval} we require four or more 3D arrays with
the same number of elements as the interpolated input diffraction
volume.  Specifically, the arrays required are the input diffraction
modulus data ($\sqrt{I(\bvec{u}})$, floating point), the current and
previous iterates ($g_n(\bvec{x})$ and $g_{n-1}(\bvec{x})$, complex
floating-point data) and the support constraint ($S$, byte data).  The
estimated modulus standard deviation, $\sigma_{\sqrt{I}}$ requires
another floating point array, but in the 3D reconstructions we set
$\sigma_{\sqrt{I}} = 0$ to reduce the memory requirement and speed up
the reconstructions.  In principle fast Fourier transforms can be
performed on arbitrary sized arrays, however it is advantageous to
perform reconstructions on a regular grid with $2^n$ grid points on a
side.  Our interpolated data array is a giga-voxel data cube
containing $1024^3 = 2^{30}$ elements which requires a total of 8\GB
of computer memory per array for single-precision complex data.  The
minimum memory footprint for single-precision iterative object
reconstruction using either the HIO or RAAR algorithm on a $1024^3$
data cube is therefore $2 \times 8\GB$ complex arrays, plus $1 \times
4\GB$ diffraction cube data and $1 \times 1\GB$ support array, giving
a total memory footprint of 21\GB of data, where we use the definition
of $1\GB=2^{30}$\thinspace bytes.  The minimum memory footprint for
performing basic HIO and RAAR reconstruction on 3D arrays of different
sizes is given in Table~\ref{tab:array-size}.  Note that this is the
minimum memory footprint needed to perform a HIO reconstruction and
that more memory may be required depending on the specific
implementation.  For example, FFT speed can be increased through use
of temporary ``workspace'' arrays which require additional memory, and
maintaining a running sum of successive images $\gamma_M$ requires an
additional complex-valued array to be retained in memory.  The memory
calculations above include only the data arrays and do not take
account of operating system requirements and the executable code
itself.

\begin{table}[htbp]
  \centering
  \begin{tabular}{ccc}
    \hline
    Array size & Single Precision & Double precision\\
    \hline
    $256^3$ & 336\MB & 592\MB\\
    $512^3$ & 2.6\GB & 4.6\GB\\
    $1024^3$ & 21\GB & 37\GB\\
    $2048^3$ & 168\GB & 296\GB\\
    \hline
  \end{tabular}
  \caption[Minimum memory requirements]{Minimum memory footprint
    required for iterative 3D phase retrieval for various array sizes. 
    The arrays required are the input diffraction data (floating point), 
    the current and previous iterates (complex single or double precision floating-point data)
    and the support constraint (byte data).}
  \label{tab:array-size}
\end{table}

The second computational challenge is efficient evaluation of the
numerous 3D Fourier transforms required for 3D phase retrieval.  The
Fourier transform of a single large data set is not trivially
parallelizable, in that the problem problem can be easily broken into
separate parallel tasks and distributed over many computer processors
as is the case, for example, with ray tracing and partially coherent
imaging where each CPU can work on a sub-set of the entire problem
without the need for intensive inter-node communication during
execution.  The nature of the Fourier transform means that any one
element of the input array affects all elements of the output,
requiring inter-node exchange of array data at each Fourier transform
step to ensure that all CPUs work together to solve the one large FFT.

We overcome the problem of efficiently calculating distributed Fourier
transforms by using the \texttt{dist\_fft} distributed giga-element
fast Fourier transform library from Apple Computer specifically
written for this project by the Apple Advanced Computation Group
\cite{Crandall:2004}.  This FFT library distributes the Fourier
transform calculation load efficiently over many processors and has
been hand-optimized to take advantage of the G5 architecture used in
the Apple Macintosh line of computers and the ``Altivec''
single-instruction-multiple-data (SIMD) floating point vector
processing unit.
Distributed FFT libraries are also available elsewhere, for example
in version 2 of the FFTW libraries \cite{FFTW05}, but at this time
these do not support SIMD vector processing extensions and proved to
be slower on our platform.  \texttt{dist\_fft} decomposes the input 3D
data set into $n_\mathrm{proc}$ discreet data slabs consisting of a $n
\times n \times (n/n_\mathrm{proc})$ voxel sub-portion of the original
data array.  Only a distinct portion of the array resides on each CPU
at any given time enabling data sets much larger than the memory of
each individual node to be computed, and the distributed memory
nature of the FFT is exploited through parallelization of all steps in
the reconstruction code.  Standard message passing interface (MPI)
\cite{MPI-url} commands are used to
communicate data between processes.

We ran fully parallelized reconstruction code on a 16-node
2.0\thinspace GHz dual-processor (32 processors total) Macintosh
Xserve G5 cluster with 4\GB RAM per node.  To maximize inter-process
communication speed we used high-speed, low-latency Mellanox
Infiniband interconnects to carry MPI traffic between compute nodes.
Using this cluster the processing time on a $512^3$ array is 2.2
seconds per iteration using the HIO phase retrieval algorithm, and an
acceptable 3D reconstruction can be produced in under 2500 iterations
for a total computation time of 2.5 hours on a $512^3$ grid.  The
individual FFT timing and total reconstruction time for typical array
sizes on this cluster is given in Table~\ref{tab:FFT-times}.

\begin{table}[htbp]
  \centering
  \begin{tabular}{ccc}
    \hline
    Array size & Time per 3D & Time per 3D \\
               & Fourier transform & reconstruction\\
    \hline
    $256^3$ & 73\ms & 10\thinspace min\\
    $512^3$ & 850\ms & 1.5\thinspace hr\\
    $1024^3$ & 7.9\s & 14\thinspace hr\\
    \hline
  \end{tabular}
  \caption[Computing times using a cluster-based Fourier transform.]{Computing times using
    a cluster-based Fourier transform and reconstruction code on 16 G5
    dual-processor Xserve compute nodes.  Fourier transform timings are wall time per
    individual FFT.   Reconstruction timings are for a complete 3D
    reconstruction consisting of 2000 iterations of HIO phase
    retrieval complete with two FFTs per 
    iteration plus other operations required to calculate the reconstruction.}
  \label{tab:FFT-times}
\end{table}

\section{Image Reconstruction}
\label{sec:reconstruction}
\subsection{Three-Dimensional Images}
\label{sec:3D-images}
A full 3D image is obtained by performing phase retrieval on the
entire 3D diffraction dataset.  The resulting volume image reveals the
structure of the object in all three dimensions and can be visualized
in many ways including forming projections through the data or slices
(tomographs) of the data. Specific segmentation analyses can be
carried out on the volume image to determine properties such as
strength of materials \cite{Ladd:1998}.  Three-dimensional
reconstructions were performed by interpolating the diffraction
intensities at $\bvec{u}_{i,j,\phi}$ onto a $1024^3$ grid.
Representations of the interpolated diffraction intensities are given
in Fig.~\ref{fig:diffraction}.  Note that the 1\degree
angular increments of the object rotation are just less than four
times larger than the 0.27\degree requirement of
Eqn.~(\ref{eq:Crowther}) for this object, and that we have a 40\degree
sector of missing data due to our limited range of object
orientations, as well as data lost to the beamstop.  The effect of the
1\degree rotation increment is apparent in
Fig.~\ref{fig:diffraction}~(b), 
where the gaps between the measured Ewald spheres are seen in the
$u_x$-$u_y$ plane (referred to as a central section) extracted from
the data cube. The
limited range of views are readily apparent in Fig.~\ref{fig:diffraction} (c),
which shows the $u_x$-$u_z$ central section.

The three-dimensional phase retrieval code described above in
Sec.~\ref{sec:computation} was applied to the assembled 3D data to
produce a full 3D reconstruction from the diffraction cube.  We
applied the Shrinkwrap algorithm, as described in
Sec.~\ref{sec:phase-retrieval}, to determine the 3D support mask and
the diffraction phases.  We performed phase retrieval using either the
real-positive real-space constraint $P_{S+}$ or the support constraint
$P_S$.  For the complex image reconstruction, as with the case of
reconstruction from central sections discussed below in
Sec.~\ref{sec:2D-projection}, the solution was regularized by first
applying the high-pass filter of Eqn.~(\ref{eq:filter}) to the
diffraction intensities.  For the real positive reconstruction the
missing amplitudes were unconstrained and were allowed to be recovered
by the algorithm.  The reconstruction success with the sparsity of
data we have in this case is undoubtedly due to the sparseness of the
object itself.  In essence the object is a membrane, and the 3D
speckles are elongated by up to 50 pixels in directions perpendicular
to the pyramid faces, as can clearly be discerned in
Fig.~\ref{fig:diffraction} (c).

\begin{figure}[htbp]
  \centering
  \includegraphics[width=0.45\textwidth]{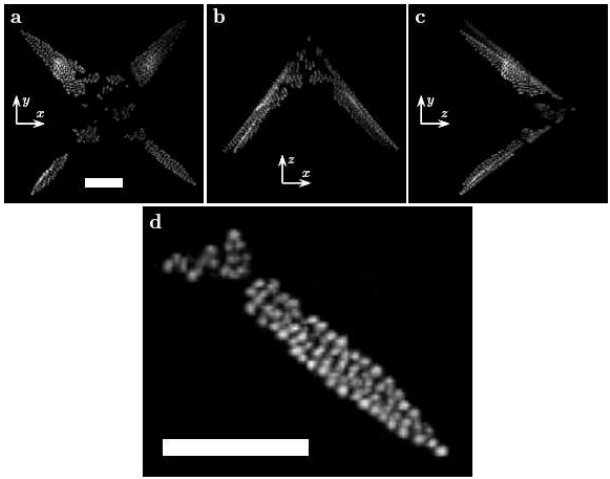}
  \caption[3D reconstruction]{Maximum value projections along three
    orthogonal directions of the reconstructed 3D image. Projections
    were performed along (a) $z$, (b) $x$, and (c) $y$ directions. (d)
    An enlarged region of (a), for comparison with
    Fig.~\ref{fig:projection}. The 3D image was reconstructed using
    reality and positivity constraints.  The scalebars are 500\nm.}
  \label{fig:3D-complex}
\end{figure}

A 3D iso-surface rendering of the real-positive constrained
reconstructed pyramid is shown in Fig.~\ref{fig:SEM}~(c), where we
also display a slice from the volume image in Fig.~\ref{fig:SEM}~(d).
Three images from the 3D pyramid image are shown in
Fig.~\ref{fig:3D-complex}, showing the maximum value projection, along
the three orthogonal axes, of the real part of the 3D image that was
reconstructed using the support constraint with real positivity,
$P_{S+}$. Each pixel of this image is given by the maximum value
encountered along the path that projects onto that pixel, and
illustrates a visualization method available only with the full 3D
image array.  In initial reconstruction trials using only a support
constraint $P_S$ we observed a linear ramp in the imaginary component.
This was essentially a phase ramp, due to a shift of the diffraction
data in Fourier space caused by an inaccurate measurement of the
location of the zero spatial frequency (the direct beam) on the CCD.
We shifted the diffraction data by an amount that minimized the
real-space phase shift, which required shifting the data by half-pixel
amounts.  This recentering of the data was necessary before we could
apply the real positive constraint on the image. Further analysis of
the images is carried out in Sec.~\ref{sec:analysis}.

\subsection{Two-Dimensional Images}
\label{sec:2D}
Two-dimensional images are useful for visualizing and quantifying
objects, and most diffraction imaging experiments performed to date
have been 2D.  However, if the object is thick, then the
interpretation of an image reconstructed from a single Ewald sphere is
not trivial.  Most notably, as compared with our full reconstructed 3D
image, the
2D image will have defocus artifacts that do not diminish in power with
displacement of the object along the beam axis. However, in some cases
obtaining a full 3D reconstruction may not be achievable, for example
when imaging non-reproducible objects with single pulses of an
XFEL. It is thus instructive to compare 2D images reconstructed from
single-view diffraction patterns with the 3D image.    

We first consider how thin an object must be to be considered two
dimensional.  In a 2D reconstruction from a single diffraction
pattern, the spatial frequencies that are passed to the diffraction
pattern are constrained on the Ewald sphere according to
\begin{equation}
  \label{eq:Ewald}
  q_z=1/\lambda - \sqrt{1/\lambda^2-q_x^2-q_y^2} \approx
  -\frac{\lambda}{2} \left ( q_x^2+q_y^2 \right ),
\end{equation}
where the approximation is for small scattering angles, or $q_x \ll
1/\lambda$.  We can define the numerical aperture of the diffraction
pattern as $\mathrm{NA} = q_{x, \mathrm{max}} \, \lambda$, in analogy
with imaging with a lens (of square aperture for the case of a square
detector, with the NA defined here along the half-width of the square
rather than the diagonal), which gives the expression of maximum
longitudinal distance of the Ewald surface, $q_{z, \mathrm{max}}
\approx -\mathrm{NA}^2/(2 \lambda)$.  For a 2D object of thickness $D
\rightarrow 0$, the 3D transform will be independent of the
longitudinal frequency $q_z$ (rods in the $z$ direction) and so the
measurement on the Ewald sphere is equivalent to measurement of the 2D
diffraction intensities $| O(q_x,q_y,0) |^2$.  In such case there will
be no artifact in the image caused by not actually measuring
intensities on $q_z = 0$.  An object of finite thickness $D$ will have
a coherent diffraction pattern with speckles of width $1/D$ in the
longitudinal direction.  If, at the highest transverse frequencies,
the Ewald sphere substantially cuts through a speckle that is centered
at $q_z = 0$, then the measurement will again be equivalent to the 2D
diffraction intensities on the $q_z = 0$ plane.  That is, we can
consider an object to be thin or two-dimensional if the Ewald
departure is no more than $1/(4D)$, or half the speckle half-width,
which corresponds to
\begin{equation}
  \label{eq:DOF}
  D < \frac{\lambda}{2 \mathrm{NA}^2},
\end{equation}
or, equivalently, the thickness $D$ must be less than a depth of
focus.  For the experiments with the pyramid object at $\lambda =
1.65\nm$ and $\mathrm{NA} = 0.084$, this thickness limit is $D =
120\nm$, which is considerably smaller than the 1.8\micron thickness
of the pyramid.

Equation (\ref{eq:DOF}) does not imply, however, that diffraction
imaging performs optical sectioning where only the parts of the object
located within the depth of focus are imaged.  The thickness limit
simply implies that the 2D single-pattern image of an object thicker
than $D$ will contain artifacts due to the information that is cut off
by the transfer function.  Consider an object containing two parts
(e.g. screens) that are separated by more than a depth of focus.  As
with coherent imaging with the equivalent aberration-free thin lens,
partial information from both screens of that object will be
transferred in the imaging process.  In fact, in diffraction imaging,
there is not necessarily any preferred image plane since, by the
Fourier shift theorem, a shift $\delta z$ of an object along the beam
axis $z$ will cause only a phase shift given by $-2 \pi \delta z \,
q_z$ and hence no change to the diffracted intensities (for small
enough $\delta z$ that the change in distance to the detector does not
change the effective NA and scale the pattern on the detector).  Note
that from Eqn.~(\ref{eq:Ewald}) the phase shifts of the 2D spatial
frequencies of the image, due to the defocus $\delta z$, will be
$\pi\,\delta z \, \lambda (q_x^2+q_y^2)$, as expected from the Fresnel
propagator \cite{Cowley:1981}.  The position of the focal plane can be
chosen in the phase retrieval step, a fact that was demonstrated
computationally and experimentally by Spence et
al. \cite{Spence:2002}.  In that work the focus of the retrieved image
of an object of two screens separated by some depth could be chosen by
setting a tight support for the features in one screen or the other.
As shown by Spence et al., once the phases of the diffraction
intensities have been retrieved, images can be generated at any
position through focus, by Fresnel propagating the image wave-field
(equivalent to applying the appropriate quadratic phase term to the
diffraction phases).

\begin{figure}[htbp]
  \centering
  \includegraphics[width=0.4\textwidth]{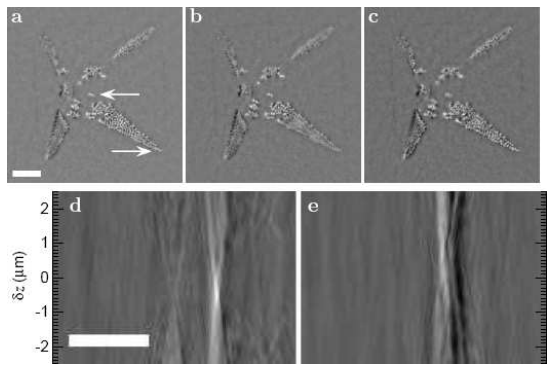}
  \caption[Defocus effects]{Real part of the image reconstructed from
    a single view diffraction pattern (a), and real part of the image
    formed by numerically
    propagating (a) by $-0.5\micron$ (b) and $+0.7\micron$ (c).  Line-outs from the image near the pyramid
    center (d) and arm extremity (e) for a range of propagation from
    -2.5\micron to +2.5\micron.  The locations of these line-outs are indicated by arrows
    in (a).  The difference of the plane of best
    focus for these two image locations is apparent. Scale bars are 500\nm.}
  \label{fig:defocus}
\end{figure}

The defocus effects of a single view are illustrated in
Fig.~\ref{fig:defocus}, where we show 2D images of the wavefield at
the pyramid object, reconstructed from a single-view diffraction
pattern. In this example, we use the diffractogram for the object
rotated by $\phi = 24\degree$ from the head-on (plan view)
orientation. The image $\gamma_M$ reconstructed by Shrinkwrap, from
the single-view diffraction pattern, is shown in
Fig.~\ref{fig:defocus} (a).  No real-space reality nor positivity
constraint was applied and the reconstructed image is complex.  For
this object and view, the edges of the object (its support) are at a
range of heights along the beam axis.  In this case the end-point
support that the Shrinkwrap algorithm arrived at was tightest around
the balls halfway along the arms of the pyramid, and consequently this
is the plane of best focus.  This focal plane gives the greatest
overall image contrast, which explains why Shrinkwrap converges to it.
The complex image can be numerically propagated, by convolution with
the Fresnel propagator, by any arbitrary amount $\delta z$.  We
generated a series of numerically refocused images, where $\delta z$
varies between $\pm 2.5\micron$, in 50\nm steps.  As the refocus
distance is increased the best focus moves along the pyramid arms to
their extremities.  The difference in focus of balls near the vertex
and arm extremities can be seen in Fig.~\ref{fig:defocus} (d) and (e)
which show $x$-$\delta z$ line-outs of the real part of the complex
image. The difference between the best focus for these two cases is
1.2\micron, which agrees with the 3D image (Sec.~\ref{sec:3D-images})
and the known geometry of the pyramid.  It should be noted that this
computational focusing does not constitute 3D imaging, but is simply
the propagation of a 2D coherent field.  The optical transfer function
(OTF) for this imaging system is the Ewald surface, and in this
situation with coherent illumination the integrated intensity of the
image does not change with defocus (a consequence of Parseval's
theorem and the invariance of the diffraction intensities with
defocus).  That is, it is unlikely that numerical defocusing of a
complicated object could give results that could be as easily
interpreted as for the pyramid-membrane test object used here.  This
situation is unlike partially-coherent imaging in a microscope, where
out-of-focus objects contribute less power to the image and some
optical sectioning can be carried out \cite{Streibl:1985}.

\begin{figure}[htbp]
  \centering
  \includegraphics[width=0.4\textwidth]{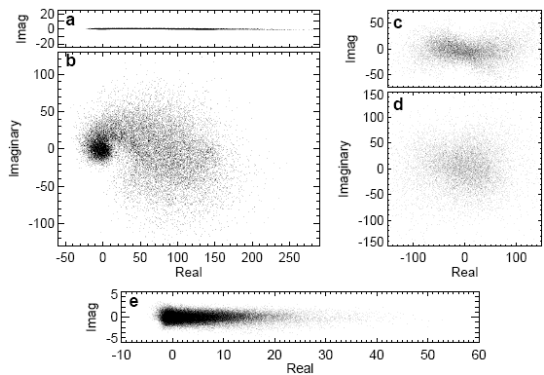}
  \caption[Argand-plane representation]{Distributions of the real-space
    complex amplitudes $\overline{\gamma}_M$, in the Argand plane, of
  simulated single-view coherent images for a 2D (a) and 3D
  (b) object consisting of 50\nm diameter gold balls, for an X-ray
  wavelength of 1.6\nm. Distributions of complex amplitudes of
  images reconstructed from experimental data,
  for (c) the infinite-depth of focus 2D projection image shown in
  Fig.~\ref{fig:projection}, (d) for the
  single-view 2D image of Fig.~\ref{fig:defocus}, and (e) the full 3D
  image. Cases (c) and (d) were reconstructed using $P_S$, and (e)
  using $P_{S+}$.} 
  \label{fig:argand}
\end{figure}

Another consequence of the ``defocus artifact'' of 2D images, is that
the 2D image of a thick real object is complex, which means that a
real-space positivity constraint cannot be applied during the phase
retrieval process.  A positivity constraint, when valid, is known to
be very effective in deriving the diffraction phases, and important in
direct methods in crystallography and a strong constraint in
diffraction imaging.  Here, a real object is one in which the object's
3D complex transmission function $o(\bvec{u})$ is real, to a
multiplicative complex constant.  Propagation of the out-of-focus
parts of the object to the selected image plane will give rise to a
large variation in the complex values of image, as demonstrated in
Fig.~\ref{fig:argand}.  Here we show the complex amplitudes of images
recovered from calculated diffraction patterns of simulated objects.
The simulated objects consisted of gold balls of equal size,
distributed in a similar way to the pyramid test object.  In the first
case (Fig.~\ref{fig:argand} a) the $z$ coordinate of all ball centers
was set to zero to construct a quasi-2D object.  Ignoring the
arbitrary phase shift, the reconstructed image is real although not
strictly positive (the negativity of the image is due to the
truncation of the diffraction pattern).  The calculated image values
are complex for the 3D object (Fig.~\ref{fig:argand} b) and there is a
rough correlation between absolute value and phase of the values.
This non-reality can also be explained by the curvature of the Ewald
sphere.  The 3D diffraction magnitudes of a real object are
centrosymmetric, whereas the Ewald sphere does not cut through both
$O(\bvec{u})$ and $O(-\bvec{u})$ \cite{Huldt:2005}.  In general, a
positivity constraint will only be applicable for the full 3D image,
2D projections (discussed in Sec.~\ref{sec:2D-projection}), 2D images
of thin objects, and 2D images of objects with a mirror-plane
symmetry.

If the object can be considered two-dimensional and positive, a positivity constraint will
have the effect of focusing the image.  Usually the support constraint is loose, and even
if the shape of the object is well known or determined from the Shrinkwrap algorithm, for
example, there may still be room for a defocused image to be contained within the support.  The degree of
defocus allowed by the support depends on how tight it is.  The defocused
image of a real 2D object with sharp edges or high frequencies will be real but include
negative values.  The focused image will be that which is most positive, and hence a
positivity constraint will select that image.  This is true not only for defocus, but for
any other low-order aberration (astigmatism, coma, etc.).  That is, without the positivity
constraint, there are slowly-varying phase modes that cannot be determined, and the number
of these modes depends on how tight the support is.  The same argument applies for 3D
reconstruction of positive 3D objects.  In this case, the phase errors will be low-order
3D modes, which cannot be thought of as focus or other aberrations of an optical system,
but are simply unconstrained phase errors in Fourier space.

\subsection{Infinite Depth-of-Focus Two-Dimensional Images}
\label{sec:2D-projection}
\begin{figure}[htbp]
  \centering
  \includegraphics[width=0.4\textwidth]{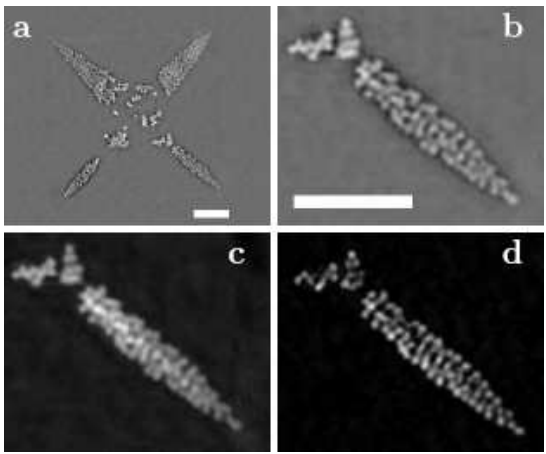}
  \caption{Infinite depth of focus projection images, for the object
    orientation $\phi = 0\degree$.  (a) Reconstruction from a 2D
    central section interpolated from the 3D diffraction dataset.  The
    reconstruction was performed without a positivity constraint,
    $E_S^2 = 0.167$.  (b)
    Enlargement of the lower right arm of (a).  (c) [and also
    Fig.~\ref{fig:SEM} (c)] Reconstruction from the 2D central section,
    using a positivity constraint, $E_S^2 = 0.072$.  (d) Projected image formed by
    integrating the full 3D reconstructed image, $E_S^2 = 0.113$.  The scalebars are
    500\nm.}
  \label{fig:projection}
\end{figure}

Defocus in a 2D image formed from a single diffraction pattern is a
consequence of the Ewald sphere OTF, as described above.  The focal
plane of the image may be varied by multiplying the Fourier transform
of the 2D image by a quadratic phase term.  In a full
three-dimensional reconstruction, there is no concept of defocus.  A
shift of the object by $\delta z$ along the beam axis causes the phase
ramp $-2 \pi \, \delta z \, u_z$ across the 3D transform.  This causes
a shift of the image, no different to shifts $\delta x$ or $\delta y$
in the other dimensions.  There is no optical axis in the
reconstructed 3D image, so there is no defocus.  Similarly, there is
no defocus in a 2D projection image formed by integrating the 3D image
along a specific direction.  A 2D projection may be recovered from the
diffraction intensities without having to first undergo a full 3D
reconstruction, and we find this is a useful step to quickly examine
our 3D datasets. By the Fourier projection theorem, the projection
image is formed from a central section in reciprocal space, e.g. the
plane $u_z = 0$ gives the projection along the $z$ axis.  We have
performed phase retrieval on central sections of the pyramid
diffraction data, by first extracting the appropriate diffraction
intensities from all views.  One example of a central section is shown
in Fig.~\ref{fig:diffraction} (b), which was generated by linear
interpolation of measured intensities at $\bvec{u}_{i,j,\phi}$ onto
the $u_z = 0$ plane.  The projection images that we reconstruct from
experimental data are superior to the reconstruction on a single Ewald
sphere.  One example is shown in Figs.~\ref{fig:projection} (a) and
(b), which can be compared with Fig.~\ref{fig:defocus}.  In the
projection images, balls at the apex of the pyramid are similar to the
balls at the base, whereas in the single view image, the balls at the
apex appear out of focus. The image of Figs.~\ref{fig:projection} (a)
and (b) was obtained using the Shrinkwrap algorithm (parameters given
in Sec.~\ref{sec:phase-retrieval}), after first regularizing by
filtering the diffraction intensities according to
Eqn.~\eqref{eq:filter}. The missing data in the arc-shaped regions
seen in Fig.~\ref{fig:diffraction} (b) were allowed to float in the
reconstruction of the complex image, according to
Eqn.~\eqref{eq:modulus}.

Figures~\ref{fig:projection} (a) and (b) depict the real part of the
complex image, and the distribution of complex values of the
reconstructed image is given in Fig.~\ref{fig:argand} (c).  As
compared to the distribution of complex values of a reconstructed
image from a single view diffraction pattern, given in
Fig.~\ref{fig:argand} (d), the values of the projection image are
clustered closer to the real axis.  This is as expected since there
are no defocus artifacts, and the object consists mostly of a single
material (gold) as was simulated in Figs.~\ref{fig:argand} (a) and
(b).  Thus, we should be able to apply the reality and positivity
constraints of Eqn.~\eqref{eq:support-positive} to the projection
image, to further improve it and allow this extra information to help
reconstruct the spatial frequency amplitudes and phases that are
missing behind the beamstop.  This was the case for
Figs.~\ref{fig:SEM} (b) and
\ref{fig:projection} (c), which is the
real-positive constrained reconstruction from the same central section
as for Fig.~\ref{fig:projection} (a).  In this case the diffraction data
were not filtered.  This reconstruction was carried out using the same
support mask derived by Shrinkwrap and used for the reconstruction of
the complex image.  Since they were constrained, the complex
amplitudes of the image were distributed along the real axis, with
some deviation from real for smaller amplitudes that could be
attributed to noise and scattering material other than gold (e.g. the silicon
nitride pyramid).  

\section{Image Analysis}
\label{sec:analysis}
Both the reconstructed X-ray 2D projection image described in
Sec.~\ref{sec:2D-projection} and the 3D image described in
Sec.~\ref{sec:3D-images} clearly show good agreement with the SEM
image of Fig.~\ref{fig:SEM} (a).  When we overlay a semi-transparent
version of the projection image of Fig.~\ref{fig:SEM} (b) on the SEM
image (a) we see that the locations of all balls visible in the SEM
match with the balls visible in the X-ray image, to within a pixel.
In the X-ray volume image however we can locate more balls than
visible in the SEM image.  The slice image of Fig.~\ref{fig:SEM} (d)
reveals that the gold balls of the object are not entirely a single
layer, but the arms of the structure are several balls deep in places.
The balls were deposited on the inside surface of the silicon nitride
pyramid, and it is clearly seen that these balls are indeed flush with
the intersecting edges of the pyramid.  The regions where the balls
are layered give rise to a higher projected image intensity which
shows up as brighter regions in the projection image of
Fig.~\ref{fig:SEM} (b).  We confirm that the 3D pyramid geometry
determined from the reconstructed volume image is consistent with the
manufacture of the pyramid.  We measure an included angle between
opposite faces of the pyramid of $70 \pm 1 \degree$, compared with the
included angle of 70.5\degree between the 111 and $1\bar{1}1$ crystal
planes of silicon.

The volume images display good spatial resolution in the $x$, $y$ and
$z$ dimensions.  Quantifying resolution is not straight forward since we do not
have an exactly known 3D standard---the SEM only
shows the surface of the object, for example, and this method cannot
reveal the 3D structure.  We estimate the resolution of our images by
examining both their Fourier and real-space representations.  In
Fourier space we base measures of resolution on the signal to noise of
measured diffraction intensities and the consistency of recovered
phases, whereas in real space we fit models to 3D images of isolated balls.

\subsection{Reconstruction Consistency and Resolution}
\label{sec:resolution}
\begin{figure}[htbp]
  \centering
  \includegraphics[width=0.4\textwidth]{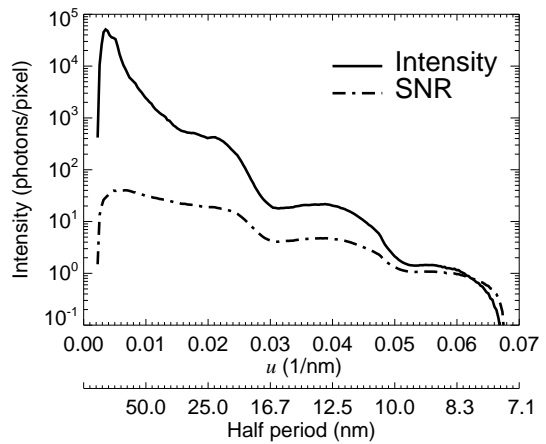}
  \caption{3D diffraction intensities $I(\bvec{u})$, averaged over
    shells of constant $u$, in units of average photon count
    per CCD pixel. The average over constant $u$ of the 3D signal to noise ratio
    (SNR) of the measured intensities is shown with a dashed line.}
  \label{fig:PSD}
\end{figure}

The performance of our imaging technique could be quantified in
Fourier space, in principle, by measuring the modulation transfer
function (MTF).  For the numerical reconstruction technique used here
this MTF would encapsulate resolution limits due to signal-to-noise,
data alignment and regions of missing data, as well as algorithm
stability and uniqueness.  The direct computation of the MTF would
require computing the ratio of the image Fourier amplitudes to the
Fourier amplitudes of the actual object, which again requires an
accurate quantitative model of the actual object structure at high
resolution.  Without such a model we can base an estimate of the upper
limit of the modulation transfer frequency cut-off on the signal to
noise of the measured diffraction data plotted in Fig.~\ref{fig:PSD}.
The largest spatial frequency used in the the interpolated 3D
diffraction dataset (recorded near the corner of the CCD) is at
$u_\mathrm{max} = \sqrt{2} N \, \Delta q = 0.068\invnm$.  At this
resolution shell we recorded an average of $< 1$ photon per pixel, and
a SNR of 1 photon per pixel at $u = 0.062\invnm$.  (Since the noise
level of our camera is considerably less than 1 photon, we assume the
noise in our diffraction patterns determined by photon shot noise.)
If we assume hypothetically that the diffraction phases are known then
the image can be fully represented, without loss of information, with
a pixel sampling of $\Delta q = 1/D$, where $D$ is the width of the
object, corresponding to $s=1$, and so we could rebin our oversampled
data into larger pixels with a correspondingly higher photon count.
Summing in this way over pixels (referred to as pixel binning) is not
the same as resampling however, and such an operation would multiply
the autocorrelation image with the Fourier transform of the summed
pixel shape, which will be a function that falls to from unity at the
image center to $2/\pi$ at the edge of the autocorrelation image.  The
effect could be deconvolved from the pattern, but we avoid that by
binning to a pixel sampling of $\Delta q = 1/(s \, D)$, with $s=2$,
which is the Nyquist critical sampling interval of the object's
autocorrelation function.  The measured data were collected at
$s=4.6$, so resampling to $s=2$ gives an average of 1 photon per pixel
(SNR = 1) at $u = 0.066\invnm$.  If we take a measure of resolution as
the frequency at which the SNR of the rebinned data is unity, then we
find that the average 3D cutoff is $0.066\invnm$ or a smallest
resolvable half period of 7.5\nm.  This is very close to the smallest
half period of 7.3\nm limited by the detector NA.

\begin{figure}[htbp]
  \centering
  \includegraphics[width=0.4\textwidth]{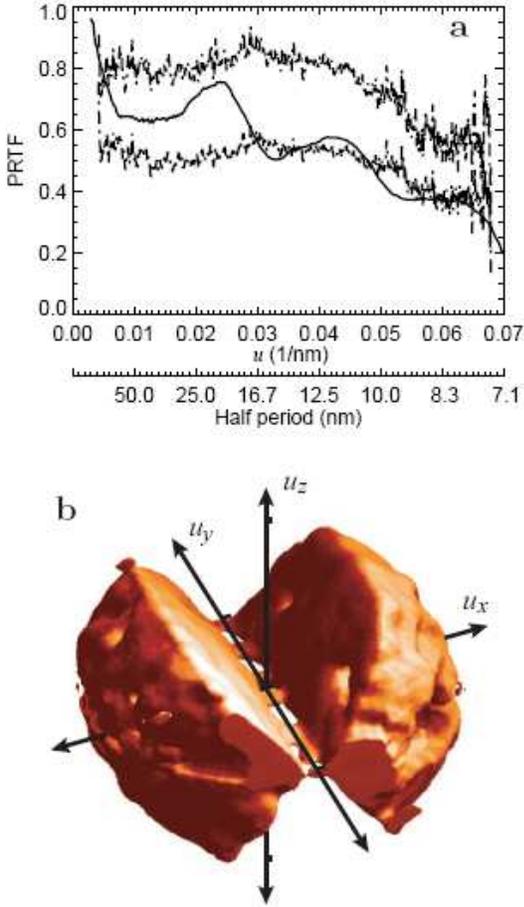}
  \caption{(Color online) (a) The phase retrieval transfer function, averaged over shells
    of constant $u$, for the real-positive 3D projection image (solid
    line) and averaged over circles of constant $u$ for the complex 2D
    image (dashed lines). The dashed line with lower values is for the
    2D projection image without correction of vortex phase modes.  (b)
    An iso-surface rendering of the 3D PRTF, at a threshold level of
    0.5. The axis tick-marks indicate 0.05\invnm.}
  \label{fig:resolution}
\end{figure}

The phase retrieval process recovers the diffraction phases with a
limited accuracy, due to factors including SNR of the diffraction amplitudes, missing
data, the inconsistency of constraints, and systematic errors in the
data (such as errors in interpolation).  These errors in phase reduce
the resolution of the synthesized image.  With a complex image a loose
support constraint will lead to unconstrained low-order aberrations, for example,
as was discussed in Sec.~\ref{sec:2D}.  In our case of reconstructing
complex 2D images, with low frequencies missing due to the beamstop,
we have observed that phase retrieval from independent random starts
may differ by a phase vortex (right or left handed), centered at the zero spatial frequency.
This too has the effect of reducing the image resolution.  One way to
quantify the effect of these phase variations is to determine the
correlation between phases retrieved from independent random starts of
the phase-retrieval algorithm.  For example, we could compute the
differential phase residual of these two solutions in the same way
that independent images are compared in cryo-electron microscopy
\cite[Chap. 3, Sec. B]{Frank:1996}.  Since we have the ability to
compute an unlimited number of reconstructions from independent random
starts, a more appropriate choice is to determine the variation in
retrieved phases as a function of resolution as suggested by V.~Elser
\cite{Shapiro:2005}.  More specifically, the average of the
independent complex reconstructions is computed, and the square of the
Fourier amplitudes of this average are compared with the measured
diffraction intensities.  Where the phases are consistently retrieved
to the same value, the squared modulus of the average will be equal to
the constrained modulus, and the ratio will be unity.  Where the
phases are random and completely uncorrelated, the average will
approach zero.  Thus, the ratio is effectively a transfer function for
the phase retrieval process, and the average image is the best
estimate of the image: spatial frequencies are weighted by the
confidence in which their phases are known \cite{Shapiro:2005}.  All
2D and 3D images displayed in this paper are averages of more than 300
independent phase retrieval trials.  That is, the best estimate of the
image is given by
\begin{equation}
  \label{eq:image-average}
\overline{\gamma}_M = \left < \gamma_M \, e^{i \phi_0} \right >,
\end{equation}
where $\left < \: \right >$ denotes an average over independent
reconstructions.  Analogous to the Modulation Transfer Function (MTF)
of a coherent imaging system, we define the Phase Retrieval Transfer
Function (PRTF) as
\begin{equation}
  \label{eq:PRTF}
  \mathrm{PRTF}(\bvec{u}) = \frac{\left | \mathcal{F}_{\bvec{u}}\left\{
      \overline{\gamma}_M\right\} \right | }{\sqrt{I(\bvec{u})}} = 
  \frac{\left | \left < \Gamma_M (\bvec{u})\, e^{i \phi_0} \right >
    \right | }{\sqrt{I(\bvec{u})}},
\end{equation}
where $\Gamma_M$ is the diffraction amplitude with retrieved phases,
the Fourier transform of Eqn.~\eqref{eq:image}.
Plots of the PRTF, averaged over shells of constant $u$ and where $I(\bvec{u})$ are non-zero, are shown in
Fig.~\ref{fig:resolution} (a) for the 3D image of
Fig.~\ref{fig:3D-complex} and for the 2D projection image of
Fig.~\ref{fig:projection} (a).

When computing the average image $\overline{\gamma}_M$, the arbitrary
multiplicative phase constant $\phi_0$ of each image must be adjusted to a common value
so that the random variation of this constant does not reduce the average, which
would result in a low value of the transfer function.  We do this for
the first reconstructed image $\gamma_M^{(0)}$ by finding the constant phase
that maximizes the real part of that image, which can be achieved by finding
the value $\phi_0$ that maximizes \cite{Fienup:1997}
\begin{equation}
  \begin{split}
    \label{eq:max-reality}
    \alpha &= \sum_{k} \Re \left\{ \gamma_M^{(0)}(k) \exp (i \phi_0) \right \}^2, \\
    &= \tfrac{1}{4} \sum_{k}  2 \left|\gamma_M^{(0)}(k)\right|^2 +
	\gamma_M^{(0)}(k)^2  e^{2i \phi_0} +\left (\gamma_M^{(0)}(k)^* \right)^2 e^{-2i \phi_0},
    \end{split}
\end{equation}
for an image with with pixels (or voxels) indexed by integers $k$ and
complex values $\gamma_M(k)$.  We maximize the square of the real part
to allow for positive and negative real values.  The value $\alpha$
can be maximized by maximizing either the second or third terms of
Eqn.~(\ref{eq:max-reality}), and we do so by finding the phase $\phi$
of the complex value $\sum_k \gamma_M^{(0)}(k)^2$, and setting $\phi_0
= -\phi/2$.  The subsequent images $\gamma_M$ are adjusted by finding
the constant phase $\phi_1$ which minimizes $\sum_{k}
|\gamma_M^{(0)}(k) - \gamma_M(k)|^2$.  This phase is that which
maximizes $\sum_{k} \Re \{ \gamma_M^{(0)}(k) \, \gamma_M(k) \,
\exp(i \phi_1) \}$, which is simply the phase of the complex value
$\sum_{k} \gamma_M^{(0)} (k)^* \, \gamma_M(k)$.

In the case of 2D images we also improve the average by separating out
the vortex modes mentioned above.  This was achieved simply by
correlating each phase retrieval solution with the previous solutions
and separating the solutions into three classes (which were found to
differ by left and right-handed phase vortices) based on the value of
the correlation.  We found that the class with the most frequent
members (60\% of trials) gave rise to the best image, wherease the
other two classes were equally frequent (20\% each) and gave rise to
images for which the balls were larger, had bright edges and reduced
intensity at their centers.  Based on the appearance of the balls we
assumed that the most frequent class was that which did not have a
vortex mode.  The effect of removing the vortex modes from the average
image is shown in Fig.~\ref{fig:resolution} (a).  As is seen in that
figure the PRTF is uniformly increased across all frequencies.  This
is due to the fact that the left and right handed vortex modes sum
together to give a variation of the modulus which varies as the cosine
of the azimuthal angle in the diffraction pattern, and which averages
to zero in the average around this complete circle for each $u$.

The resolution cutoff of the phase retrieval process can be given by
the spatial frequency at which the PRTF extrapolates to zero.  For all
cases here, this frequency is greater than the measurement cutoff of
$u_\mathrm{max} = 0.068\invnm$, or resolution of 7.4\nm.  A more
conservative estimate of the resolution is given by the frequency at
which the PRTF reaches a value of 0.5.  For the vortex-corrected 2D
reconstruction this occurs just at $u_\mathrm{max}$, but for the 3D
image this corresponds to 0.048\invnm, or a resolution of 10.4\nm.  In
this case the average resolution cutoff is worse than the 2D case
because the 3D PRTF is diminished along the $u_z$ direction where the
diffraction data are missing (which reduces the average over the $u$
shell).  This is illustrated in Fig.~\ref{fig:resolution} (b), where
we display the 3D PRTF as a surface for which it has a value of 0.5.
The PRTF is not defined in the regions of missing diffraction data,
which are seen as the missing wedges in the surface.  It is seen that
the resolution is approximately the same in all directions of
$\bvec{u}$ where intensities were measured.

When applied to the average image $\overline{\gamma}_M$, the modulus
constraint error $E_M^2$ of Eqn.~\eqref{eq:modulus-error} is equal to
the intensity-weighted integral over $\bvec{u}$ of
$|1-\mathrm{PRTF}(\bvec{u})|^2$.  That is, it gives a single measure
of how well diffraction intensities of the average image agree with
the measurement.  This is generally higher than the metric $E_M^2$
applied to the iterate $g_n$, which gives an estimate for how well the
algorithm fits the intensity data.  The value of $E_M^2$ applied to
the average 3D image is 0.368, and 0.059 for the average 2D projection
image that was corrected for vortex phase errors (0.312 without vortex
correction).  We expect that a similar correction of low-order phase
modes in the 3D image would lead to a similar improvement in the error
metric, and the relatively high value of $E_M^2$ for the average 3D
image is due to the overall filtering due to the variation of these
low-order phase modes.
 
We can also compute the agreement of the average image
$\overline{\gamma}_M$ to the real-space support constraint $E_S^2$ of
Eqn.~\eqref{eq:error}.  We find a value of 0.228 when applied to the
average 3D image and 0.167 for the average 2D complex-valued
projection image reconstructed from the central section.  Note however
that in the 3D image the support $S$ accounts for 0.10\% of the image
voxels whereas $S$ covers 4.1\% of the pixels in the projection 2D
images, and so the average error per pixel outside the support is much
less for the 3D that the 2D reconstruction.  We find with the addition
of the real-space positivity constraint that $E_S^2$ of the average 2D
projection image decreases from 0.167 to 0.072.  However, in this case
the modulus constraint error $E_M^2$ increases from 0.059 to 0.172.


\subsection {Real-space resolution}
\label{sec:real-space}
\begin{figure}[htbp]
  \centering
  \includegraphics[width=0.45\textwidth]{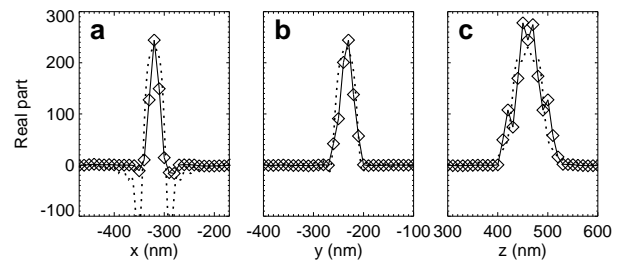}
  \caption{Line-outs of the real part of the reconstructed complex
  amplitude 3D image, for three orthogonal directions (a) $x$, (b)
  $y$, and (c) $z$, through the isolated single ball at the pyramid
  apex.  Coordinates are relative to the center of the 3D image
  array. Dashed lines show lineouts from a simulated 3D coherent image
  with a cube OTF with a 60\degree missing sector. }
  \label{fig:lineouts}
\end{figure}

The measures of resolution from the SNR and PRTF reveal the effects of
noise, consistency of the diffraction data, and how well the image
obeys the imposed constraints.  These measures are contributors to the
overall image resolution.  A direct measure of a lower limit of
resolution can be obtained by examining the images of isolated and
closely spaced gold balls.  Line-outs of the isolated ball located on
the lower left arm of Fig.~\ref{fig:3D-complex} (a) are shown in
Fig.~\ref{fig:lineouts}, for all three orthogonal directions.  The
ball image has full widths at half maximum (FWHM) of 30, 35, and 70\nm
in the $x$, $y$, and $z$ directions, respectively.  Images of other
isolated balls in the object are very similar to that shown in
Fig.~\ref{fig:lineouts}.  Assuming the balls are 50\nm in diameter, we
obtain a good fit to the images by modeling a coherent imaging system
with an optical transfer function (OTF) that is unity within a cube of
half-width $0.05\invnm$ (centered at the zero frequency) and which is
zero within a sector of 60\degree as rotated about the $y$ axis, and
centered about the $z$ axis.  Line-outs of the modeled coherent
images, computed by convolving an isolated 50\nm sphere with the
Fourier transform of the OTF (that is, the point spread function, or
PSF), are shown as dashed lines in Fig.~\ref{fig:lineouts}.  The FWHM
of the modeled coherent image are 36\nm, 40\nm and 64\nm in the $x$,
$y$, and $z$ directions, respectively, in good agreement with the
reconstructed image. We do not expect the model to be an exact fit to
the data, since the actual PSF is more complicated and depends on the
details of the phase retrieval, which is better characterized by the
PRTF in Fig.~\ref{fig:resolution}.  However, the fits are reasonable
and the widths of the modeled PSF are in good agreement with the
measures of resolution obtained from analysis of the diffraction
intensities and recovered phases.  The modeled point spread function
(PSF), given by the Fourier transform of the OTF, has a half-width of
$10\nm \times 10\nm \times 40\nm$.  Here the half width is defined as
the distance from the central maximum of the PSF to the first zero.
Since the imaging process is coherent, the image width depends on the
phase of the PSF, which has a different distribution for the $x$ and
$y$ directions.  This explains the variation of image widths in the
$x$ and $y$ directions, and why the image FWHM in these directions are
in fact smaller than the ball diameter.  As expected, the resolution
in the $z$ direction is much worse than in the $x$ and $y$ directions,
due to the missing sector of data that arises from recording
diffraction over a limited range of angles.

\section {Summary and Conclusions}
We have presented 3D images reconstructed \emph{ab initio} from
coherent X-ray diffraction, which exhibit high resolution in all three
dimensions.  These images are the highest resolution 3D images of
non-periodic objects where the resolution is comparable in all three
dimensions.  The work presented here marks an important advance in
that we have fully demonstrated the experimental methods to collect 3D
coherent X-ray diffraction and the computational tools to manage the
data, and reconstruct 3D images of more than $10^9$ voxels.

The coherent X-ray diffraction recorded from our 3D test object
comprised of 140 views, at 1\degree intervals, and extend to a maximum
spatial frequency of 0.068\invnm, or a smallest reconstructible
half-period of 7.4\nm.  Although we cannot exactly quantify the
resolution of the image, which would require knowing the object's 3D
structure, we have determined the consistency of the retrieved phases
which gives us an estimate of an upper bound of the MTF of the imaging
process.  Our analysis shows we can consistently retrieve phases out
to the maximum spatial frequency recorded.  This consistency measure
does not tell us anything about systematic errors, such as
interpolation of the data, errors in assigning spatial frequency
$\bvec{u}$ to the intensities (imperfect knowledge of the beam
center), and missing data due to the beam stop or limited range of
object orientations.  However, we easily resolve 50\nm spheres that
are touching each other, and from such image line-outs, and
comparisons of reconstructed X-ray images with the SEM image, we have
confidence that our achieved image resolution is close to our upper
estimate.

We have found that our Shrinkwrap algorithm \cite{Marchesini:2003},
which determines the object support \emph{ab initio}, is robust and
works well even with missing Fourier-space data due to limited object
orientations or the beamstop.  The phase retrieval process can be
essentially characterized by a 3D MTF (the Phase Retrieval Transfer
Function, or PRTF) which is influenced by the noise of the measured
diffraction intensities.  While the algorithm lets the amplitudes at
the locations of missing data to also be recovered, these values are
not consistently reconstructed and are averaged to zero, leaving worse
resolution in the depth ($z$) direction.  We expect that with a
dataset collected over the full range of sample orientation angles we
would achieve equal resolution in all three dimensions.  As it is, we
obtained an estimate of 10 nm in $x$ and $y$ and 50 nm in $z$.

We have shown that high-NA X-ray coherent imaging of thick objects can only
properly be carried out in the context of three dimensions. Here we
define high-NA imaging of thick objects to be imaging under conditions that
lead to a depth of focus less than the depth of the object, in any of
its orientations.  Since the imaging is coherent, a 2D image of a thick
object in any one view will exhibit defocus artifacts which do not
diminish in overall power with the degree of defocus and which lead to
difficulties in the interpretation of the image.  In addition, these
artifacts cause the image of a real positive object, for example, to
be complex, hence hampering quantitative evaluation of the image.
Two-dimensional images free of defocus artifacts can be quickly
generated from central sections extracted from the diffraction data.
Three-dimensional images are synthesized from the entire 3D
diffraction dataset.  The tools are now in place to perform full 3D
reconstructions of thick samples.  Currently we have reconstructed
arrays with almost $2\times 10^9$ elements.  If the minimum
oversampling of $\sqrt[3]{2}$ relative to Bragg sampling is used in
each dimension, then this would correspond to objects of width
9.5\micron at a pixel spacing of 10\nm, or a resolution of 7\nm along
the diagonal.  When single-particle XFEL imaging at atomic resolution
becomes feasible, then these demonstrated computational capabilities
could be used to reconstruct objects of 480\nm width at 0.7\nm
resolution, for example.  This would correspond to a large virus, or a
large protein complex such as the ribosome.

\section*{Acknowledgments}
\label{sec:acknowledgements}

We wish to thank Ray Mariella (LLNL) for the idea of using a silicon
nitride pyramid as a test object, and Jackie Crawford and Dino Ciarlo
(LLNL) for determining its manufacturing process.  We thank Janos Kirz
(LBNL and Stony Brook) for technical advice and extensive discussions
about our experiments.  We acknowledge stimulating discussions with
Abraham Sz\"{o}ke (LLNL), G\"{o}sta Huldt (U. Uppsala), and Eugene
Ingerman (CBST).  We gratefully acknowledge Richard Crandall and the
Advanced Computations Group (Apple Computer, Inc.) for the development
of the \texttt{dist\_fft} software.  This work was performed under the
auspices of the U.S. Department of Energy by University of California,
Lawrence Livermore National Laboratory under Contract W-7405-Eng-48.
This work has been supported by funding from the National Science
Foundation. The Center for Biophotonics, an NSF Science and Technology
Center, is managed by the University of California, Davis, under
Cooperative Agreement No. PHY 0120999.


\end{document}